\documentclass[twocolumn]{aastex631}

\hypersetup{linkcolor=blue,citecolor=blue,filecolor=cyan,urlcolor=blue}
\usepackage{xcolor}
\usepackage[encapsulated]{CJK}

\newcommand{\reopt}{$R_{\rm eff., opt}$}
\newcommand{\logrkpc}{$\log\ R_{\rm eff., opt}$ (kpc)}

\newcommand{\logm}{$\log M_*$}
\newcommand{\logmsol}{$\log M_*/M_\odot$}

\received{\today}
\revised{XXX}
\accepted{YYY}

\submitjournal{ApJ}

\shorttitle{JWST UNCOVER/MegaScience: The Growth of Galaxy Sizes at $4< z <8$}
\shortauthors{Miller et al.}

\graphicspath{{./}{paper_figs/}{paper_figs/app_examples/}}

\begin{document}

\title{JWST UNCOVERs the Optical Size - Stellar Mass Relation at $4<z<8$: Rapid Growth in the Sizes of Low Mass Galaxies in the First Billion Years of the Universe}

%% Main author
\author[0000-0001-8367-6265]{Tim B. Miller}
\affiliation{Center for Interdisciplinary Exploration and Research in Astrophysics (CIERA), Northwestern University, 1800 Sherman Ave, Evanston, IL 60201, USA}

%%Morphologogeurs
\author[0000-0002-1714-1905]{Katherine A. Suess}
\affiliation{Department for Astrophysical \& Planetary Science, University of Colorado, Boulder, CO 80309, USA}

\author[0000-0003-4075-7393]{David J. Setton}\thanks{Brinson Prize Fellow}
\affiliation{Department of Astrophysical Sciences, Princeton University, Princeton, NJ 08544, USA}

\author[0000-0002-0108-4176]{Sedona H. Price}
\affiliation{Department of Physics and Astronomy and PITT PACC, University of Pittsburgh, Pittsburgh, PA 15260, USA}

%%Builders
\author[0000-0002-2057-5376]{Ivo Labbe}
\affiliation{Centre for Astrophysics and Supercomputing, Swinburne University of Technology, Melbourne, VIC 3122, Australia}
\author[0000-0001-5063-8254]{Rachel Bezanson}
\affiliation{Department of Physics and Astronomy and PITT PACC, University of Pittsburgh, Pittsburgh, PA 15260, USA}

\author[0000-0003-2680-005X]{Gabriel Brammer}
\affiliation{Cosmic Dawn Center (DAWN), Niels Bohr Institute, University of Copenhagen, Jagtvej 128, K{\o}benhavn N, DK-2200, Denmark}

\author[0000-0002-7031-2865]{Sam E. Cutler}
\affiliation{Department of Astronomy, University of Massachusetts, Amherst, MA 01003, USA}

\author[0000-0001-6278-032X]{Lukas J. Furtak}\affiliation{Physics Department, Ben-Gurion University of the Negev, P.O. Box 653, Be’er-Sheva 84105, Israel}

\author[0000-0001-6755-1315]{Joel Leja}
\affiliation{Department of Astronomy \& Astrophysics, The Pennsylvania State University, University Park, PA 16802, USA}
\affiliation{Institute for Computational \& Data Sciences, The Pennsylvania State University, University Park, PA 16802, USA}
\affiliation{Institute for Gravitation and the Cosmos, The Pennsylvania State University, University Park, PA 16802, USA}

\author[0000-0002-9651-5716]{Richard Pan}\affiliation{Department of Physics and Astronomy, Tufts University, 574 Boston Ave., Medford, MA 02155, USA}

\author[0000-0001-9269-5046]{Bingjie Wang (\begin{CJK*}{UTF8}{gbsn}王冰洁\ignorespacesafterend\end{CJK*})}
\affiliation{Department of Astronomy \& Astrophysics, The Pennsylvania State University, University Park, PA 16802, USA}
\affiliation{Institute for Computational \& Data Sciences, The Pennsylvania State University, University Park, PA 16802, USA}
\affiliation{Institute for Gravitation and the Cosmos, The Pennsylvania State University, University Park, PA 16802, USA}

\author[0000-0003-1614-196X]{John R. Weaver}
\affiliation{Department of Astronomy, University of Massachusetts, Amherst, MA 01003, USA}

\author[0000-0001-7160-3632]{Katherine E. Whitaker}
\affiliation{Department of Astronomy, University of Massachusetts, Amherst, MA 01003, USA}
\affiliation{Cosmic Dawn Center (DAWN), Denmark} 

%% UNCOVER Collaboration
\author[0000-0001-8460-1564]{Pratika Dayal}
\affiliation{Kapteyn Astronomical Institute, University of Groningen, P.O. Box 800, 9700 AV Groningen, The Netherlands}

\author[0000-0002-2380-9801]{Anna de Graaff}
\affiliation{Max-Planck-Institut f\"ur Astronomie, K\"onigstuhl 17, D-69117, Heidelberg, Germany}

\author[0000-0002-1109-1919]{Robert Feldmann}
\affiliation{Department of Astrophysics, Universität Zürich, Winterthurerstrasse 190, 8057 Zürich, Switzerland}

\author[0000-0002-5612-3427]{Jenny E. Greene}
\affiliation{Department of Astrophysical Sciences, Princeton University, 4 Ivy Lane, Princeton, NJ 08544, USA}

\author[0000-0001-7201-5066]{S. Fujimoto}\altaffiliation{Hubble Fellow}
\affiliation{
Department of Astronomy, The University of Texas at Austin, Austin, TX 78712, USA
}
\author[0000-0003-0695-4414]{Michael V. Maseda}
\affiliation{Department of Astronomy, University of Wisconsin-Madison, 475 N. Charter St., Madison, WI 53706, USA}

\author[0000-0003-2804-0648 ]{Themiya Nanayakkara}
\affiliation{Centre for Astrophysics and Supercomputing, Swinburne University of Technology, PO Box 218, Hawthorn, VIC 3122, Australia}

\author[0000-0002-7524-374X]{Erica J. Nelson}
\affiliation{Department for Astrophysical and Planetary Science, University of Colorado, Boulder, CO 80309, USA}

\author[0000-0002-8282-9888]{Pieter van Dokkum}
\affiliation{Astronomy Department, Yale University, 52 Hillhouse Ave,
New Haven, CT 06511, USA}

\author[0000-0002-0350-4488]{Adi Zitrin}
\affiliation{Department of Physics, Ben-Gurion University of the Negev, P.O. Box 653, Be'er-Sheva 84105, Israel}

\begin{abstract}
We study the rest-frame optical and ultraviolet morphology of galaxies in the first billion years of the Universe. Using JWST data from the UNCOVER and MegaScience surveys targeting the lensing cluster Abell 2744 we present multi-band morphological measurements for a sample of 995 galaxies selected using 20-band NIRCam photometry and 35 using NIRSpec Prism spectroscopy over the redshift range of $4<z<8$. The wavelength-dependent morphology is measured using \texttt{pysersic} by simultaneously modeling the images in 6 NIRCam wide filters covering the rest-frame UV to optical. The joint modeling technique increases the precision of measured radii by 50\%. Galaxies in our sample show a wide range of  Sersic indices, with no systematic difference between optical and UV morphology. We model the size-mass relation in a Bayesian manner using a continuity model to directly fit the redshift evolution while accounting for observational uncertainties. We find the average size of galaxies at $\log M_*/M_\odot=8.5$ grows rapidly, from 400 pc at $z=8$ to 830 pc at $z=4$. This is faster evolution than expected from power law scalings of the Hubble parameter or scale factor that describe well previous results at $z<2$. This suggests that different and/or much stronger processes affect low mass systems during the epoch of reionization. The measured logarithmic slope (0.25) and scatter (0.23 dex) are non-evolving. We discuss the remarkable consistency of the slope and scatter over cosmic time in the context of the galaxy-halo connection.
\end{abstract}

\keywords{Galaxy Formation (595), High-redshift Galaxies (734), Galaxy radii (617), Scaling Relations (2031) }

\section{Introduction} \label{sec:intro}

The physical processes that control the growth and evolution of galaxies also determine their spatial extent or size. Therefore, by studying the distribution of sizes for a population of galaxies we can learn about their formation histories. The size of a galaxy is often parameterized using the half light radius, $R_{\rm eff}$, which encloses half of the total light, most commonly measured at rest-frame optical wavelengths i.e. 5000\AA. Most galaxies are bright at rest-optical wavelengths and they are readily accessible from ground-based observations, making the rest-optical ideal for the local universe. 

In particular it is common to study the relationship between \reopt\ and total stellar mass, the so-called size-mass relationship. Growth mechanisms, such as galaxy mergers and star-formation, affect the size and mass in different ways. Therefore the distribution of galaxies in the size-mass plane and its evolution with redshift can be used to learn about the physical processes which affect galaxies. Observationally the \reopt -mass relationship has been well characterized from the local Universe out to cosmic noon using a combination of ground \citep{shen2003,trujillo2006, lange2015, kawinwanichakij2021} and space-based \citep{vanderwel2014,Allen2017, mowla2019,nedkova2021} imaging. Additionally, it represents an important constraint on cosmological galaxy formation simulations \citep{furlong2017,genel2018,pillepich2018}. 

In this paper we focus on star-forming (late-type) galaxies. For this population a consistent picture of the size-mass relationship has emerged: \reopt{} is positively correlated with \logm{}. Save for the most massive galaxies (\logmsol$\gtrsim10.5$; \citealt{mowla2019b}), the size-mass relationship is observed to follow a power-law relationship with a slope of $\sim 0.2$. This slope is observed to remain consistent over many orders of magnitude in stellar mass \citep{lange2015, Carlsten2021, nedkova2021} and constant with redshift out to $z\approx 2$ \citep{vanderwel2014, mowla2019, kawinwanichakij2021}. The distribution around this relation is observed to be log-normal, where the width is observed to be roughly $\sigma \sim 0.2$, and again, consistent across the entire mass and redshift range studied. The normalization of this relation, reflecting the average size of galaxies, is observed to decrease with redshift, such that at a given mass galaxies are smaller at high redshift.

%% Talk about how this is interpreted maybe?
%It has long been theorized that the sizes of star-forming galaxies are intimately connected to their halos. The disk formulation, postulates that during formation that the gas accreeted onto galaxies conserve (some fixed fraction) of the halo's angular momentum. Thus the size of the disk can be directly related to the properties of the halo, such as the mass and  angular momentum, or spin. \citep{Fall1980,mo1998}. inside out growth, simulation results etc

The reddest wavelength coverage of the \textit{Hubble Space Telescope} (HST), approximately 1.6 $\mu m$, limited the ability to study the rest-frame optical emission beyond a redshift of 2. At higher redshifts, the rest-frame ultraviolet (UV) had to be used instead. UV emission from galaxies is acutely sensitive to dust absorption and recent star-formation. Therefore optical and UV morphologies can differ greatly and the rest-UV is not a reliable tracer of total stellar mass~\citep{ma2015,marshall2022}. While many studies have investigated the UV morphology of the earliest galaxies~\citep{Kawamata2018,shibuya2021,Bouwens2022}, it cannot be directly compared to optical sizes. Before the \textit{James Webb Space Telescope} (JWST), the rest-optical size-mass relationship in the early universe was unknown.

The longer wavelength coverage of the JWST provides high-resolution imaging out to $\sim 5 \mu m$, allowing the study of rest-frame optical emission from galaxies out to $z=8$, into the epoch of reionization. This has allowed the extension of the study of optical morphology out to higher redshifts \citep{Kartaltepe2023,Sun2023, Morishita2024, ormerod2024, Allen2024}. 

In this study we will focus on the rest-frame optical and UV morphology of star-forming galaxies at $4<z<8$ in the field of the lensing cluster Abell 2744. Specifically we will focus on data from two surveys: Ultradeep NIRSpec and NIRCam Observations before the Epoch of Reionization \citep[UNCOVER;][]{Bezanson2022} as well as the Medium-bands Mega Science \citep[MegaScience;][]{Suess2024}, which provide photometric coverage using all 20 NIRCam bands along with spectroscopic followup. The deep observations in the broad filters along with the spectral coverage of the medium band filters provide unparalleled photometric redshift accuracy \citep{naidu2024} and stellar mass precision.  We focus our analysis on the low mass galaxies (\logmsol$ \lesssim 9.5$) where the small area ($\sim30$ arcmin$^2$) field provides a representative and statistical sampling of the population. 

The galaxies at this epoch are inherently faint. Even with the impressive sensitivity of JWST NIRCam and the deep observations of UNCOVER and MegaScience, aided by gravitational lensing, they are often just above the detection threshold. Given the signal-to-noise threshold for morphological analysis is higher than that of detection~\citep{vanderwel2012}, some of the resulting measurements will have large uncertainties. In order to maximize the precision of the inferred parameters and accurately capture the uncertainties we develop a joint multi-band fitting technique to the Bayesian inference framework of \texttt{pysersic} \citep{pasha2023}. Joint fitting of multiple bands, connected by a smooth function, allows the information from all of the images to be used simultaneously for inference, maximizing the constraints on the morphological parameters of these faint galaxies. We combine the morphological constraints, along with photo-z, stellar mass and lensing constraints in order to infer the parameters of the size-mass relationship.

The rest of the paper is organized as follows: Section~\ref{sec:sample} introduces the data and galaxy sample used, the multiband fitting approach, and its implementation in \texttt{pysersic}, and application to our galaxy sample is shown in Section~\ref{sec:sizes}. The structure of galaxies, comparing the rest-optical to rest-UV is analyzed in Section~\ref{sec:structure}, the inference procedure for fitting the size-mass relationship and the accompanying results are shown in Section~\ref{sec:size_mass}. These results are discussed in Section~\ref{sec:disc} including connecting the growth of galaxies across cosmic time and comparison to predictions from simulations, ending with the conclusions in Section~\ref{sec:conc}. Throughout the paper we adopt a cosmology following the 9-year results from WMAP \citep{Hinshaw2013}. All magnitudes are reported in the AB system \citep{Oke1983}. We assume a Chabrier initial mass function~\citep{chabrier2003}. 

\section{Data and Galaxy Sample}
\label{sec:sample}
\subsection{Data and Sample Selection}
\label{sec:samp}
This study is based on JWST data primarily from the UNCOVER \citep{Bezanson2022} and MegaScience \citep{Suess2024} surveys. The raw JWST data presented in this article were obtained from the Mikulski Archive for Space Telescopes (MAST) at the Space Telescope Science Institute. The specific observations analyzed can be accessed via \dataset[doi: 10.17909/7yvw-xn77]{https://doi.org/10.17909/7yvw-xn77}. Both target the lensing cluster Abell 2744 at $z=0.307$, which contains one of largest areas of  moderate magnification, making it an excellent target for deep observations to study the earliest galaxies. The combination of UNCOVER and MegaScience image the same $\sim30$ arcmin$^2$ spanning the entire NIRCam filter suite; including 8 broad band and 12 medium band filters to depths of 28-30 mag AB. For this study we target galaxies galaxies at $4<z<8$. For this range we can measure both restframe optical and UV morphology from NIRCam alone. For our galaxy sample we combine a spectroscopic and photometric selection, to benefit from both while overcoming some of the challenges; the photometric selection ensures a complete sample that is not biased by the spectroscopic selection function while we can use the galaxies which have spectroscopic redshifts to provide stricter constraints.

Our sample is built off of the UNCOVER Data release 3 photometric catalog, originally presented in \citet{Weaver2024}, and updated to include the MegaScience data in \citet{Suess2024}.  In short, the catalog is created using \texttt{aperpy}\footnote{Available here: https://github.com/astrowhit/aperpy} and is selected on a noise-equalized F277W+F356W+F444W image. For our analysis we further restrict the selection to ${\rm SNR}_{\rm F277W+F356W+F444W} > 10$. We also make a cut on lensing magnification at $\mu < 4$, using the \texttt{v2} lensing map \citep{Furtak2023}. This is to avoid any of the highly magnified arcs. In appendix~\ref{sec:app_size_mass_res} we test whether an even stricter cut on magnification, $\mu < 1.5$, affects the results presented in this paper and find that it does not.

To select galaxies we use the catalog of stellar population properties derived using the MegaScience observations and described in \citet{Suess2024} following the methodology in \citet{Wang2024}. Galaxy properties are inferred using \texttt{Prospector} \citep{leja2017,johnson2021} using the informative, redshift-dependent \texttt{Prospector}-$\beta$ priors \citep{Wang2023}. The photometric redshifts show good agreement when compared to spectroscopic redshifts with an outlier fraction of about 7\%.~\citep{Suess2024}. Galaxies are selected using the following cuts:

\begin{itemize}
    \item \texttt{use\_phot}$ = 1$
    \item $n_{\rm bands} > 6$ and $\chi^2 / n_{\rm bands} < 5$
    \item $4 < z_{50} < 8$ and $z_{16} > 2$
    \item $\mu < 4$
\end{itemize}

This makes up the ``photometric'' samples used throughout the paper. In addition we discard any galaxies that fall under the LRD selection defined in \citet{Labbe2023} as they may be AGN and their stellar masses are highly uncertain \citep[e.g.][]{wang2024b}. We additionally select based on the spectroscopic observations of the UNCOVER survey.  For that sample we select targets from the spectroscopic catalog presented in \citet{Price2024} using the criteria:

\begin{itemize}
    \item $4 < \texttt{z\_spec} < 8$
    \item \texttt{flag\_zspec\_qual} $\geq 2$
    \item \texttt{flag\_successful\_spectrum} $= 1$
    \item $\mu < 4$
\end{itemize}

These cuts are used to restrict this sample to galaxies with successful reductions and robust spectroscopic redshifts. The  $\texttt{flag\_zspec\_qual}$ cut ensures at least one robust, or two marginally, detected spectral features were used to determine the redshift~\citep[For a full description see]{Price2024}. We additionally discard any galaxy with known broad lines indicating they are an AGN~\citep{Greene2024} This makes up the ``spectroscopic'' sample used in this study. We note that the spectroscopic selection supersedes the photometric selection, so no galaxy is used twice. Stellar masses for galaxies in the spectroscopic sample are calculated using the same \texttt{Prospector}-$\beta$ framework with the redshift restricted to $\pm 0.05$ away from the spectroscopic value, in Data Release 4~\citep{Price2024}.

To calculate the mass-completeness of our sample we follow the method presented in \citet{Pozzetti2010}. From our sample we take the lowest 30\% of stellar masses (median of the posterior) and rescale them such that their magnitude in the F277W+F356W+F444W detection image matches the detection limit, which we assume to be 30~\citep{Weaver2024}. We then take the 95th percentile of the distribution of rescaled masses as the mass limit. This results in a mass limit, accounting for magnification, across the entire redshift range of $\log M_*/M_\odot = 7.8$. After applying this stellar mass cut, the sample of spectroscopically selected galaxies contains 45  galaxies and the sample of photometrically selected galaxies contains 1306 galaxies.

\section{Joint Modeling of Morphology in Multiple Bands}
\label{sec:sizes}
\begin{figure}
    \centering
    \includegraphics[width = \columnwidth]{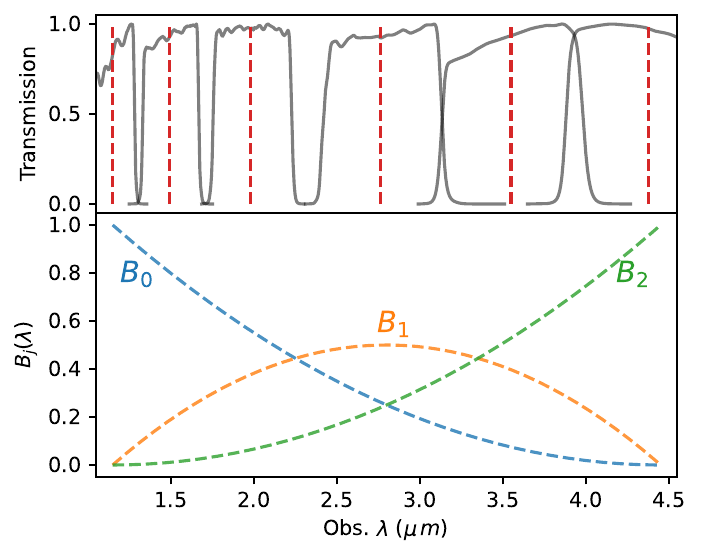}
    \caption{\textit{Top:} Transmission curves for the six filters used in the multi-band fitting procedure are shown along with the effective wavelength of each. \textit{Bottom:} The basis functions used in the B-spline used to model the $r_{\rm eff}$, Sersic index and axis ratio variations as function of wavelength.}
    \label{fig:basis_filters}
\end{figure}

\subsection{Implementation in \texttt{pysersic}}
In order to better constrain the morphological measurements we model all of the broad band UNCOVER images simultaneously. The underlying assumption is that the morphology, i.e. effective radius or Sersic index, varies smoothly with wavelength. One consideration is the presence of high equivalent width lines that dominate the light in a specific filter. In Appendix \ref{sec:app_examps} we look at the morphology of galaxies where this high EW lines are likely to dominate the flux and find it has little effect on the measured morphology~\citep[e.g.][]{Zhu2024}. By modeling multiple images simultaneously using a low-order parametrization to characterize the wavelength variation we hope to increase the constraining power. This gain arises from both decreasing the number of free parameters, as compared to fitting each band individually, along with increasing the number of independent constraints by simultaneously modeling images in multiple bands.

To perform this multi-band modeling we extend the \texttt{pysersic} software package with the \texttt{multiband} module~\citep{pasha2023}.\footnote{First implemented in \texttt{pysersic} version 0.1.4, see https://pysersic.readthedocs.io/en/latest/multiband-example.html for an example} We build off of the rendering and Bayesian inference framework already built in \texttt{pysersic}, and extend the capabilities to fit multiple bands simultaneously. This implementation, described below, is heavily inspired by that of \texttt{galfitm} \citep{Haussler2013}.

Each galaxy is modeled using a single Sersic profile. During the fitting procedure, the central position in both coordinates and position angle are assumed to be the same across the multiple bands and therefore fit as one parameter each. The total flux in each band is allowed to vary independently. The effective radius, Sersic index, and ellipticity are parameterized to vary smoothly with wavelength. For this we use a basis spline or B-Spline. This method uses a set of pre-computed basis functions $B_i$ which are then modified by coefficients, $c_i$ to produce the resulting wavelength dependence following:
\begin{equation}
    S(\lambda) = \sum_i^N c_i B_i (\lambda)
\end{equation}
Here $S(\lambda)$ represents the morphological quantity of interest. Since the basis functions, $B_i$ are pre-computed, the coefficients are the free parameters that control the wavelength dependence of our morphological parameters. For this work we use a 2nd order B-spline with 3 knots spanning the entire wavelength range, as shown in Figure~\ref{fig:basis_filters}. The B-spline curves are derived using \texttt{scipy} \citep{scipy}. Three knots provides enough flexibility to capture large scale variations in the morphology over the wavelength range that are expected. We did not find any systematic difference in the measured parameters when we tested higher order spines with more knots but found they produced un-physical oscillations in the morphological parameters as a function of wavelength.

These splines are flexible, low-order representations of the wavelength dependence much like low order polynomials that have been used previously. B-splines offer several benefits over polynomials used in previous implementations like \texttt{galfitm} \citep{Haussler2013}. They are less sensitive to extrapolation issues and ringing near the boundaries fit, known as Runge's phenomenon. Additionally since the splines are local, i.e. only affected by the filters in their vicinity rather than a polynomial fit to all filters, they are less sensitive to the extent of the wavelength range used. Even as the range of restframe wavelengths probed shifts with redshift we do not expect the inferred parameters to vary systematically.
Most importantly it is much easier to control the limits of the resulting function. In our case there are natural limits to the parameters, e.g. ellipticity must be between 0 and 1, which are difficult to enforce when fitting a polynomial. Given the normalization of B-splines, if we simply enforce these limits on the coefficients, the resulting function will also be bounded by 0 and 1 across the whole wavelength range.

\subsection{Application to UNCOVER/MegaScience}
In this study we fit the morphology of galaxies using the six wide NIRCam filters taken as part of the UNCOVER survey (F115W, F150W, F200W, F277W, F356W and F444W), which span the observed wavelength range of $1.15\mu m - 4.44 \mu m$. These bands are the deepest exposures representing the best morphology constraints and least sensitive to contamination from emission lines. The transmission curves and effective wavelengths for each filter are highlighted in Figure~\ref{fig:basis_filters}. Each filter is fit on the same $0.04$ arcsec/pixel scale, this allows for a simpler implementation of the multi-bands fitting algorithm since all of the images can be aligned on the same spatial grid. We note that this is not the native scale for the SW filters. For each filter we retrieve cutouts of the BCG-subtracted image, weight map (with Poisson Noise included), and empirical PSFs. See \citet{Weaver2024} for more information on the data products. For each galaxy we simultaneously fit neighbouring sources that are within 1 arcsec and brighter than one magnitude fainter than the target galaxy in the LW detection image. All other sources are masked, utilizing the segmentation map provided by \citet{Weaver2024}. Since that is based on a F277W+F356W+F444W detected catalog, we found unmasked sources in the shorter wavelength filters that are not present in the source catalog. We derive an additional mask using a stack of the SW filters (F115W, F150W and F200W) using similar \texttt{sep} detection parameters, that is applied only to these filters. For each cutout we subtract the median value of the unmasked pixels but do not model a sky background.

For each galaxy the total number of free morphological parameters is 18: The central position and position angle, which are independent of wavelength, the total flux in each of the 6 filters and three B-spline coefficients for each of $r_{\rm eff.}$, $n$ and ellipticity. We adopt a 1 pixel wide Gaussian prior centered on the position in the \citet{Weaver2024} photometric catalog for the central position and a flat prior on position angle (between 0-$\pi$) The flux prior is set to be a Gaussian with the mean given by the total flux in each filter and the width given by a third of that flux and truncated at 0. Priors for the spline coefficients, and thus the resultant parameters, for $r_{\rm eff.}$, $n$ and ellipticity are set to be uniform between 0.25-25 pixels, 0.65-4 and 0-0.9, respectively.

When neighboring galaxies are jointly modeled, they are approximated by single Sersic profiles modeled separately in each individual band. For these neighboring sources we set informed priors on the parameters based on a multivariate Gaussian approximation to the posterior. This estimate is obtained using Laplace's approximation at the Maximum a-posteriori point and then widened by a factor of two for the prior. This approach of using data-informed priors, often referred to as empirical Bayes, can lead to falsely over-confident results since you are using the data twice, to set the prior and do inference \citep[see][]{gelman2013}. In our case we only use data-informed priors for the neighboring sources where our primary goal is to simply to remove their light in order to better fit the primary galaxy. Morphological parameters of the neighbors are not used in the analysis and these priors greatly increase the sampling speed and efficiency for the galaxy of interest.

We find the bulk of the unmasked ``sky'' pixels are well described by a Gaussian centered at zero. However, a small fraction, $\sim 10^{-2.5}$ are strong outliers that are better described by a wider Cauchy distribution with a scale of 2 times the median pixel rms. This fraction depends strongly on sky position as well as filter; therefore we fit an additional free parameter to account for these pixel outliers, $\log f_{\rm Cauchy}$, that is allowed to vary between  $-5$ to $-1.5$. The loss function is represented by a mixture of these the standard Gaussian function and this additional Cauchy distribution~\citep[similar to what is described in][]{hogg2010}.

Inference is performed using \texttt{numpyro}~\citep{bingham2019,phan2019} probabilistic programming language using the automatic differentiation capabilities of \texttt{jax}~\citep{jax2018}. Before sampling we re-parameterize the variables using a multivariate normal approximation of the prior following \citet{hoffman2019}. We then draw samples from the posterior using the No U-turn sampler~\citep{hoffman2014, phan2019} using four chains with 750 warm-up and 1500 sampling steps each. 

\begin{figure}
    \centering
    \includegraphics[width = \columnwidth]{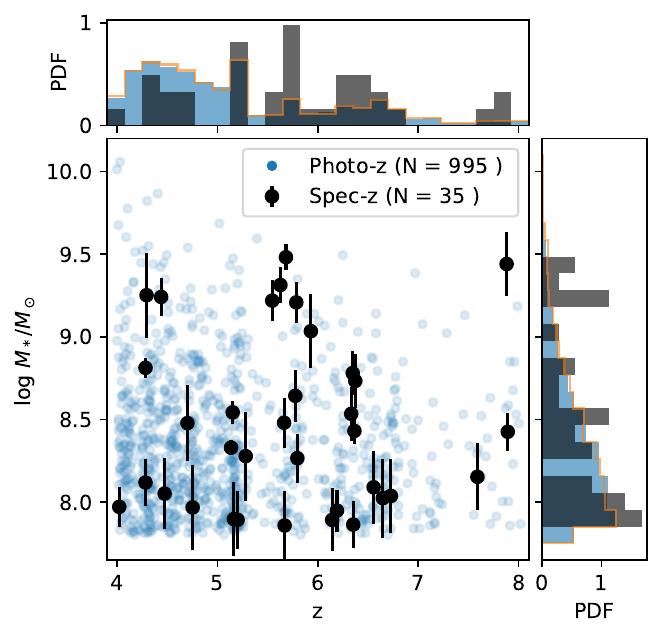}
    \caption{The distribution of mass and redshift for the spectroscopically selected (black) and photometrically selected (blue) galaxy sample in this analysis. These samples are those with well measured sizes, see Sec.~\ref{sec:sizes}, which represents about 85\% of the original sample based on the selection described in Sec.~\ref{sec:samp}. The distribution of the original sample is shown in an orange line on the histograms showcasing that there are no systematic differences.}
    \label{fig:mass_z_sample}
\end{figure}

\subsection{Assessing the Multi-band Morphological Fits}

Once sampling is complete we perform a series of quality checks. To ensure robust sampling we discard galaxies where the $\hat{r}$ metric, which compares inter- to intra- chain variance \citep{vehtari2021}, is $<1.05$ or where the effective sample size, using both the bulk and tail estimates, is $>500$. This removes roughly 9\% of the samples. We then limit the analysis to well behaved fits by cutting all galaxies where the mean flux in any of the bands is greater than 2 magnitudes (or a factor of 5) discrepant from the catalog value and where the $\chi^2$ per pixel within 20 pixels of the center is greater than four. We find that this cutoff effectively eliminates galaxies for which the fit is bad,  e.g. the galaxy is in a crowded region but retains complex sources for which the Sersic profile is still a reasonable representation of the overall light profile. In Appendix~\ref{sec:app_examps} we show examples in different $\chi^2$ slices. This further reduces the sample by 4\%. To ensure we do not include any point sources in our sample, where we cannot reliably infer the morphology, we also cut galaxies where the 5\% percentile of the $r_{\rm eff}$ posterior is less than 0.5 pixels at either rest-frame optical or UV wavelengths (using spectroscopic or the median photometric redshift), removing another 2\% of the sample. This leads to a sample of 995 photometrically selected galaxies, and 35 spectroscopically selected galaxies that are used for the analysis presented in this paper. Figure \ref{fig:mass_z_sample} displays the distribution of this sample of galaxies in the stellar mass - redshift plane. We also show the distribution of the original sample and find there are no major systematic differences between the original and analysis sample.

\begin{figure*}
    \centering
    \includegraphics[width = \textwidth]{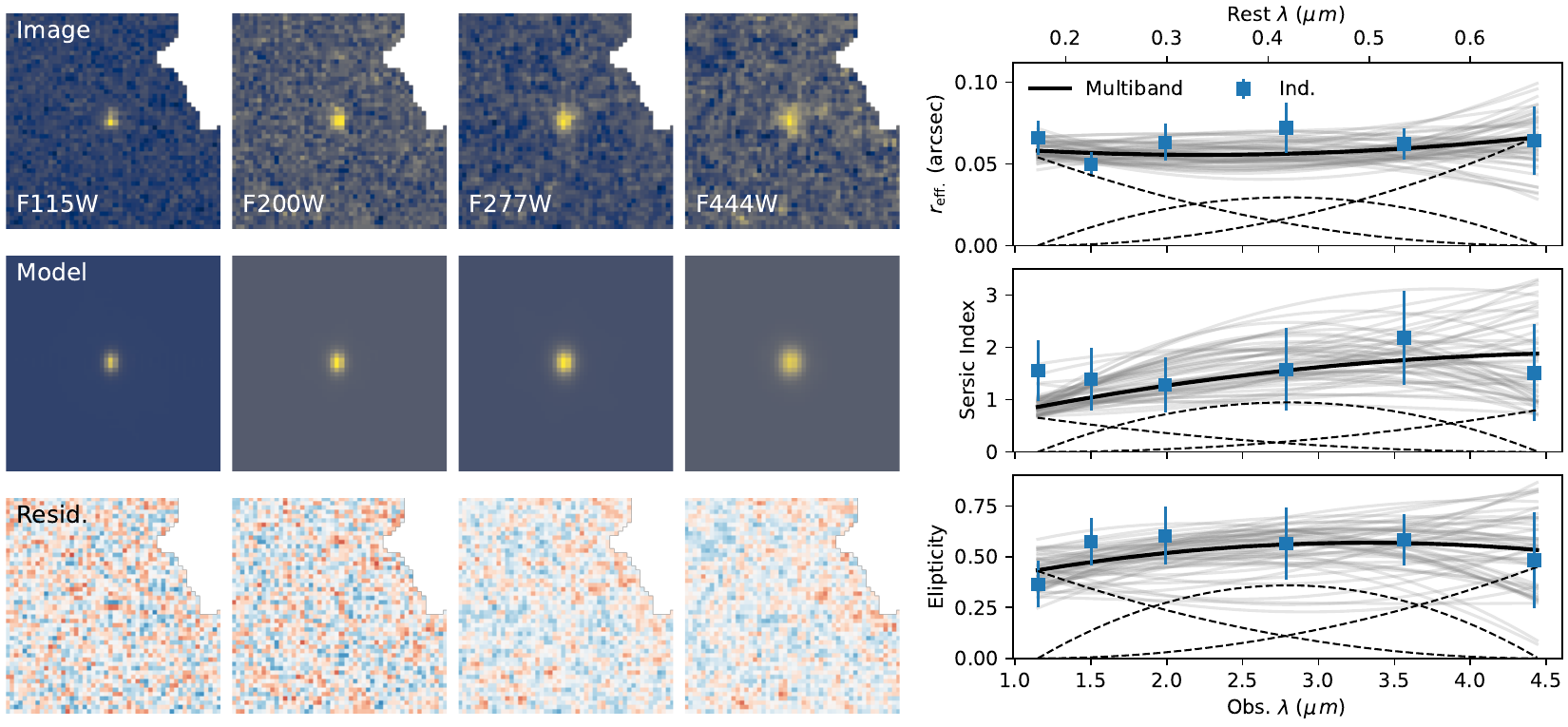}
    \caption{A detailed look at a typical fit, showcasing the galaxy with MSA ID 34495 in UNCOVER spectroscopic release at a measured redshift of 5.67. \textit{Left:} Observed and model images are shown using an arcsinh stretch for four filters spanning the wavelength range used. The single Sersic profile fits well with very little structure in the residuals. \textit{Right:} The recovered size, Sersic index and ellipticity for the multiband fits alongside the corresponding fits to individual bands shown in blue. The median of the posterior for the multiband fit is shown in a black line with the individual basis splines shown in dashed lines. Fifty random draws from the posterior distribution for each parameter are displayed as grey lines to help visualize the measurement uncertainties on the morphological parameters. The multiband fits show smoother variation with wavelength and tighter constraints compared to the individual band fits.}
    \label{fig:multi_showcase}
\end{figure*}

\begin{figure*}
    \centering
   \includegraphics[width = 0.49\textwidth]{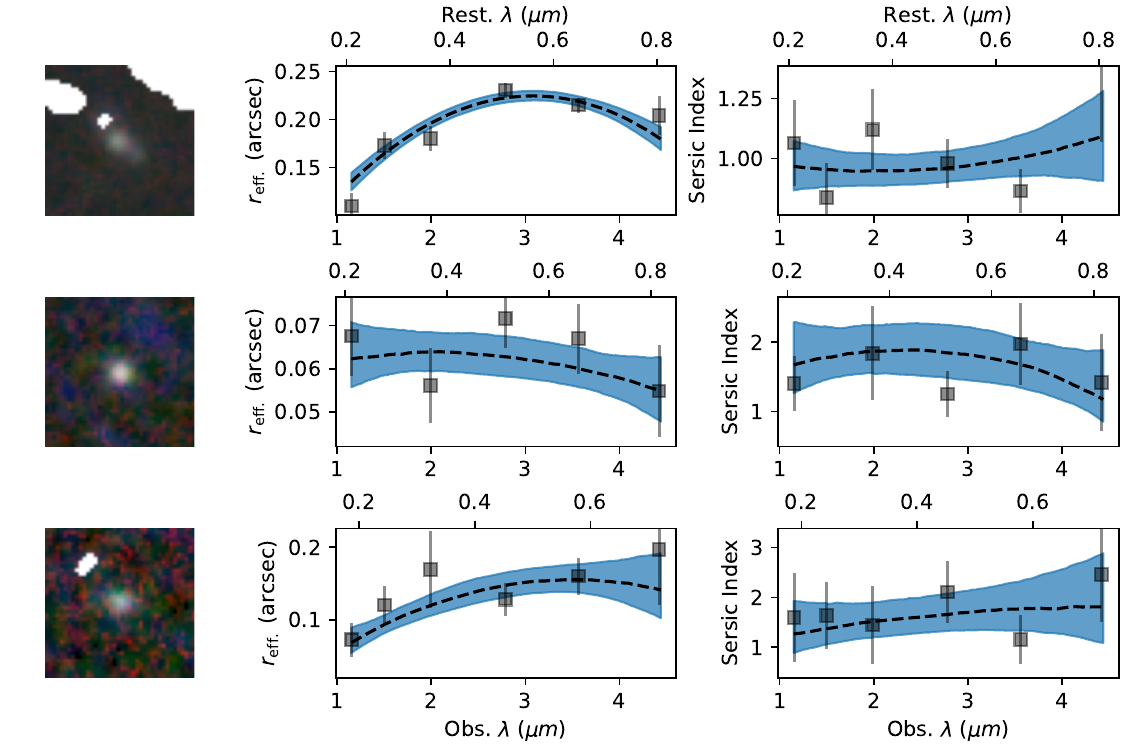}
    \includegraphics[width = 0.49\textwidth]{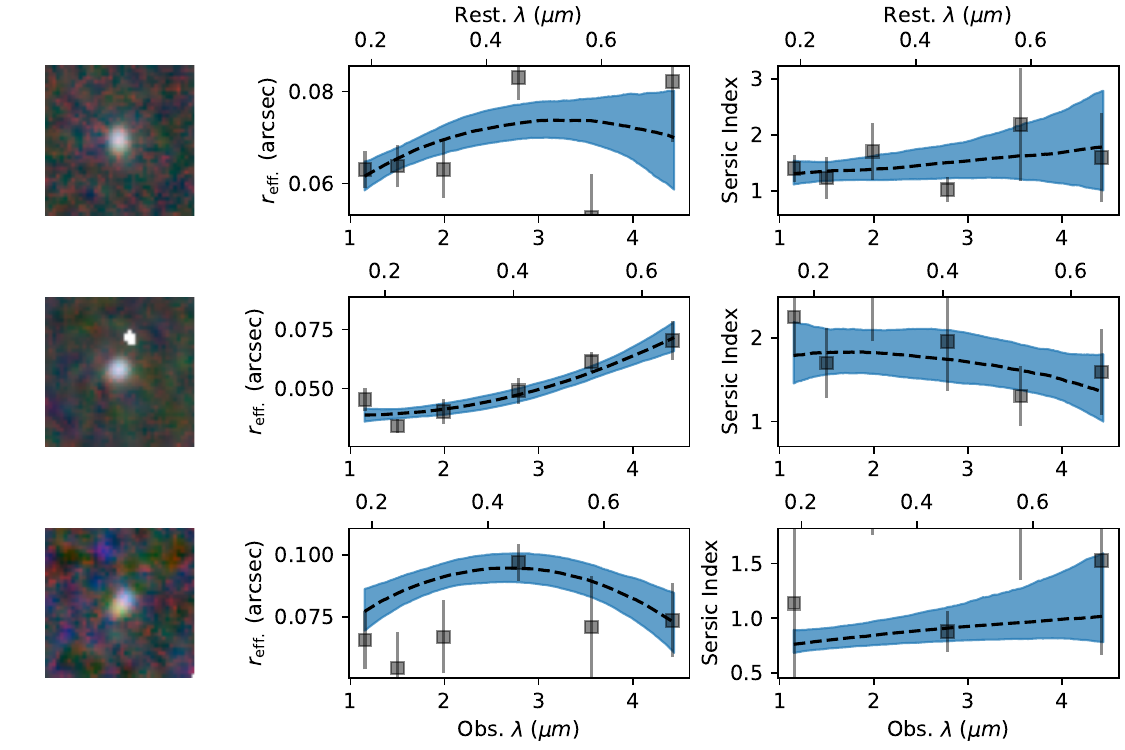}
    \caption{The result of the joint multiband fitting procedure applied to six representative example galaxies. Cutouts show the PSF-matched F150W, F277W and F444W filters. The wavelength dependent effective radius and Sersic index are shown. The dashed line displays the median while the shaded blue region shows the 16\%-84\% interval. For comparison the results of morphological fits to individual bands are shown as grey points. There is a variety of morphology displayed with no systematic wavelength-dependent trend.
    }
    \label{fig:examp_gals}
\end{figure*}

A detailed look at a multiband fit for an example galaxy is shown in Figure~\ref{fig:multi_showcase}. Images, model images from the Maximum-a-posteriori (MAP) values and residuals are shown for four of the six filters used. Beside these images we show the measured $r_{\rm eff.}$, $n$ and ellipticity as function of wavelength, showcasing the difference between the multiband and individual band fits. The wavelength dependent morphology for fifty random draws from the posterior are included to help visualize the uncertainties on measured parameters. The B-spline model employed in the multiband fit ensures smooth variation with wavelength while still accounting for overall trends. Multiband fits to several more example galaxies are shown in Figure~\ref{fig:examp_gals}. The recovered effective radius and Sersic index as a function of wavelengths are plotted alongside an RGB image of each galaxy created using the F444W,F277W and F150W filters respectively. 

There is a large diversity of morphologies displayed. These galaxies range from disk-like, with $n \approx 1$, to much steeper profiles, with $n>3$. Additionally there is a variety of wavelength dependencies, the two galaxies on the top row show radii that are consistent across the entire wavelength range while on the bottom row the two galaxies show strong gradients in the morphology with radii increasing at longer wavelengths. Unlike what is observed in local galaxies ~\citep[e.g.][]{kelvin2012} the dependence of $r_{\rm eff.}$ and $n$ on wavelength does not always appear to be monotonic. The same behavior is present in the individual band fits so this does not appear to be an artifact of the multi-band fitting procedure. The physical cause of this is unclear. One possibility is a central starburst with off-center dust attenuation or photometric contamination by strong emission lines. The latter can lead to one band dominating the signal to noise and therefore warping the fit. 

An example of this can be seen in the lower galaxy in the right column of Fig~\ref{fig:multi_showcase} where the F277W band is much higher S/N and therefore dominates the multiband fit. In this scenario the smooth spline assumed in the multiband fitting is not necessarily a good approximation of the individual band fits. However in other cases, as seen in Appendix~\ref{sec:app_examps} the galaxy is high enough S/N in the surrounding bands that the spline used in the multiband fit is able to successfully interpolate the smooth behavior even with the presence of an "outlier" band. To investigate the frequency of this behavior we analyze the distribution of fractional uncertainty on the measure $r_{\rm eff}$ for the six individual band fits for each galaxy. We compare the minimum fractional uncertainty to the median of the six bands to find galaxies where the ratio of minimum to median is $<<1$ indicating that one band dominates the signal to noise. The ratio of minimum to median fractional uncertainty is less than 0.25 for less than $<1\%$ of galaxies and less than 0.5 for $25\%$ of galaxies in our analysis sample. This suggests this behavior where one band dominates the S/N is quite rare and should not affect our overall analysis.

\begin{figure}
    \centering
    \includegraphics[width=0.95\columnwidth]{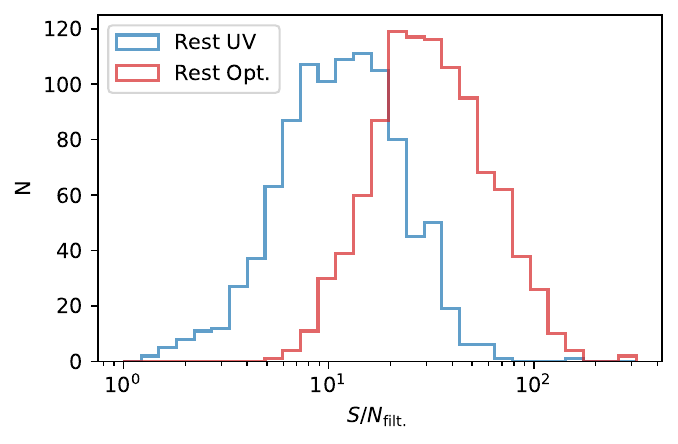}
    \caption{The distribution of single filter SNR for galaxies in our analysis sample, considering the filter closest to restframe UV and optical, 2,000\AA\ and 5,000\AA, respectively. Considering the rest optical filter, 89\% of galaxies in our sample have SNR$>15$, a limit traditionally considered acceptable for Sersic profile fitting.}
    \label{fig:snr_dist}
\end{figure}

To better place our sample in context of previous studies, specifically those using single band morphology measurements, in Figure~\ref{fig:snr_dist} we show the distribution of SNRs in then single filter which corresponds closest to restframe UV (2,000 \AA) and restframe optical (5,000 \AA). This corresponds to F277W and F115W at $4<z<5.5$, F356W and F150W at $5.5<z<7$ and F444W and F200W at $7<z<8$ for rest optical and rest UV, respectively. For rest optical emission we find that $87\%$ of the galaxies in our sample have SNR$>15$, which has been previously suggested as a benchmark for when reliable Sersic parameters can be measured.~\citep{vanderwel2012,ono2013, Shibuya2015} Additionally all galaxies have $SNR_{\rm opt.}>5$. For rest-UV the distribution is shifter lower, only $35\%$ of sources have $SNR_{\rm UV}>15$ and $89\%$ have $SNR_{\rm UV}>5$.

To assess if we are able to accurately measure morphological parameters in this low SNR regime we perform a set of recovery tests, outlined in Appendix~\ref{sec:recov_tests}. In short we select a set of high $SNR_{\rm opt.}$ ($>30$) galaxies from the analysis sample and add noise to degrade the images until they reach target $SNR_{\rm opt.}$ of 5 and 10. We then apply the same morphological fitting procedure and assess how well the originally measured parameters were recovered. We find that $R_{\rm eff.}$ can be recovered well at the low SNR values, to better that $5\%$ and $10\%$ on average for $SNR_{\rm opt.}$ of 10 and 5 respectively. Comparing the residuals between recovered and input parameters to the measured uncertainties we find close to a unit normal distribution, except for a preference for recovered sizes to be small for a fraction of galaxies, indicating the uncertainties are largely representative and well calibrated. Recovery for the index $n$ is not consistent. At these low SNRs the posterior returned largely resembles the prior giving essentially no constraints. To account for the inability to accurately constrain $n$, we limit our sample to SNR$>15$ to ensure reliable measurements.

%%% Some more example galaxies, not sure we need to show these?

\begin{figure}
    \centering
    \includegraphics[width = 0.8\columnwidth]{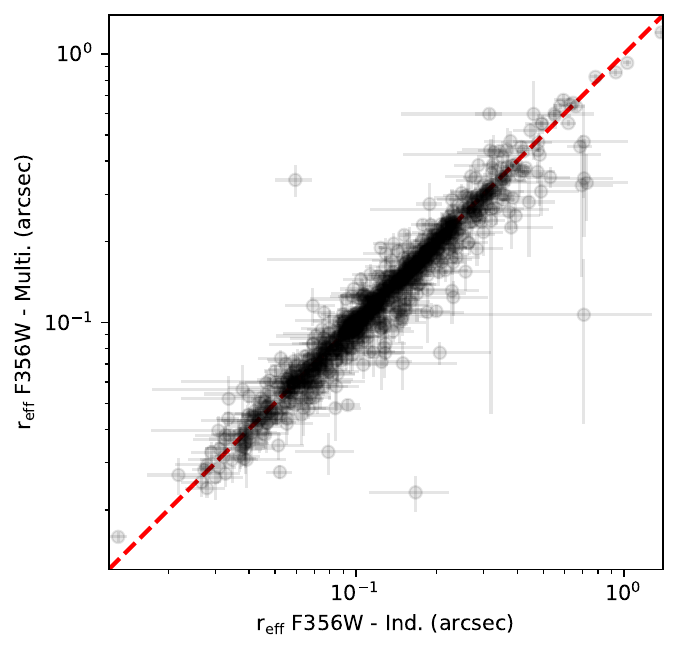}
    \includegraphics[width = \columnwidth]{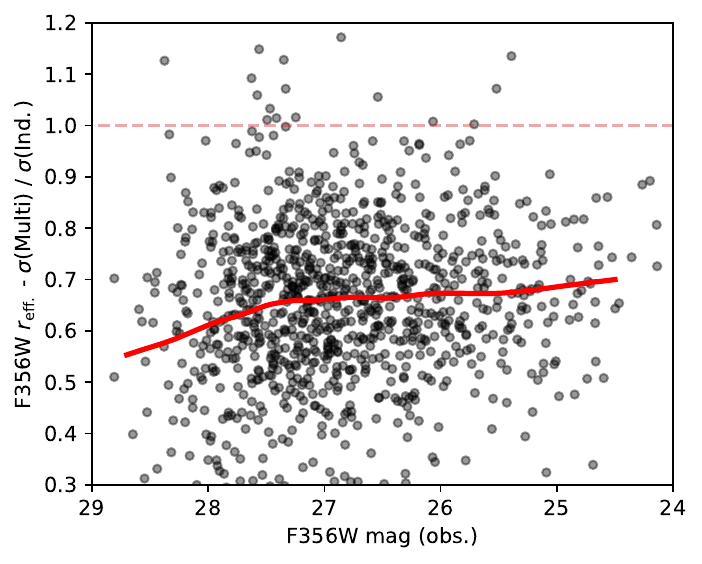}
    \caption{Comparing the multiband morphological fits to individual fits in F356W. \textit{Top:} The effective radius measured from individual band morphological fits compared to the multi-band fit, with the red-dotted line showcasing the 1-to-1 relation. We find excellent agreement with the median difference $<2\%$.  \textit{Bottom:} The ratio between the 1-$\sigma$ width of the F356W $r_{\rm eff}$ posterior comparing the multi-band fitting procedure vs. modeling the band individually. This ratio is plotted as a function of the observed F356W magnitude with the \texttt{loess} locally weighted regression shown as a red line. The ratio for almost all galaxies is less than one with a median of 0.67, indicating that on average the multiband fitting technique produces $50\%$ tighter constraints on average compared to fitting each band separately.}
    \label{fig:multi_ind_comp}
\end{figure}

We focus on the increased constraining power when performing the multiband fits in Figure~\ref{fig:multi_ind_comp}. The 1-sigma width, measured as half of the 16\%-84\% interval, is calculated for the effective radius measured in F356W, using both the multiband fitting technique along with fitting the F356W band individually. The ratio of the posterior widths are plotting as a function of the observed F356W magnitude, with the red line displaying the \texttt{loess} measured average~\citep{cleveland1979,cappellari2013}. This ratio is lower than one for almost all galaxies, with an overall median of 0.67, meaning the width of the multiband $r_{\rm eff}$ posterior is consistently lower compared to an individual band. Both of the extremes, where the uncertainty ratio is $>1$ or $<0.3$, tend to be galaxies with large band-to-band variations in their individually measured morphology, for the former the F356W constraints are tight and broad for the latter. On average, we are achieving 50\% tighter constraints using the multiband fitting technique. There is a brightness dependence to the ratio of posterior widths, the median for the fainter sources, $m_{\rm F356W} > 28 $, is 0.58, while for the brighter sources, with $m_{\rm F356W} < 26$, the median is higher, at 0.68. The improvement in constraints is more pronounced for fainter galaxies but is still relevant across the entire populations. 

\section{The Rest-frame Optical and UV structure of galaxies at $4<z<8$}
\label{sec:structure}

\begin{figure}
    \centering
    \includegraphics[width=\columnwidth]{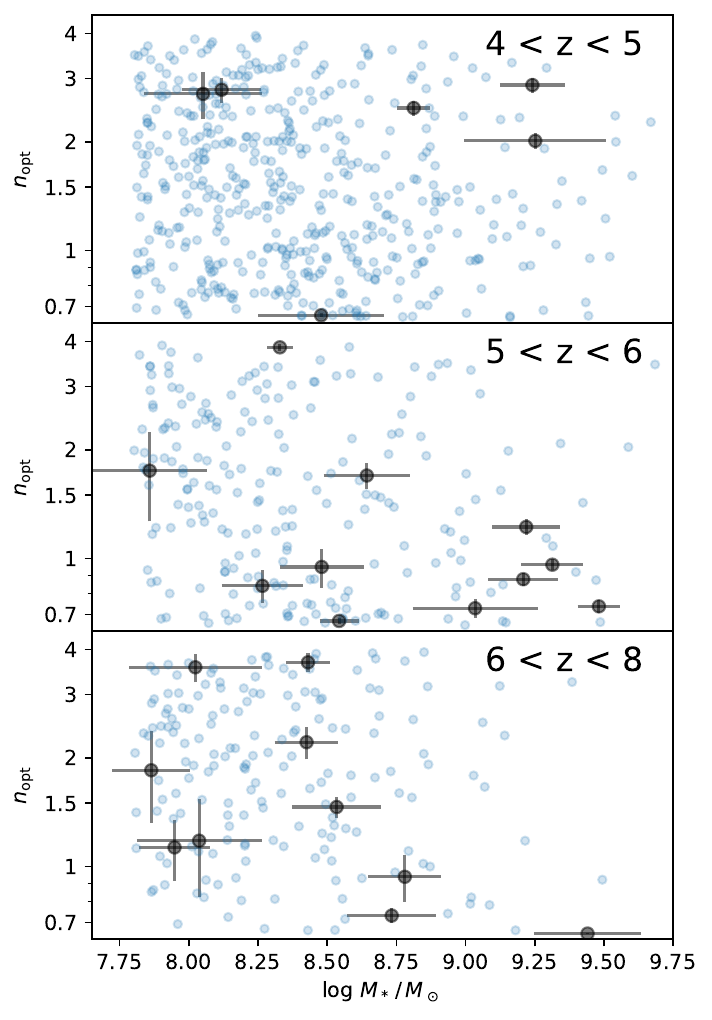}
    \caption{The distribution of our sample in the optical Sersic index and stellar mass plane split into three redshift slices. We note that the vertical axis is shown in a logarithmic scale. Black points with error bars show the spectroscopic sample while the lighter blue points represent the photometric sample. There is no discernible trend with stellar mass but we find the prevalence of galaxies with exponential disk profiles, with measured $n_{\rm opt} < 1.5$, increases at lower redshifts.}
    \label{fig:mass_index_opt}
\end{figure}

To investigate the evolution of the morphological structure of galaxies in our sample, in Figure~\ref{fig:mass_index_opt} we analyze the distribution of the Sersic index measured at optical wavelengths, $\lambda_{\rm rest} = 5000$\AA, ($n_{\rm opt}$) and how it evolves with redshift. In our parameter recovery tests (See Appendix~\ref{sec:recov_tests} we find that the Sersic index cannot be reliably measured for low SNR sources, so for this analysis we have limited our sample to that with $SNR_{\rm opt.} > 15$. These cuts may bias the sample shown in this figure, particularly towards high masses but are necessary to ensure robust morphological measurements. In all figures in this section the spectroscopic sample is shown in black with individual error bars and the median of the posterior for the photometrically selected galaxies in light blue. We wish to highlight that individual Sersic index measurements are often quite uncertain as evidenced by the significant error bars on the symbols from the spectroscopic sample. The other caveat is that we have not applied any correction to the measured index to account for gravitational lensing. Since we are in the low magnification regime ($\mu < 4$) there are likely only minor effects on the profile but for detailed studies these effects need to be investigated further.

We find a wide range of Sersic indices across all of the masses and redshifts considered in this sample. There is no clear trend in Sersic index with stellar mass; galaxies of all masses span the entire allowed range from 0.65-4. Studying the evolution with redshift, we find an increase in the prevalence of low index profiles at $z<5$. We consider the fraction of profiles that have exponential disk-like profiles, which we define as having the posterior median $n_{\rm opt} < 1.5$. We find the disk-like fraction at $4<z<5$ to be $52 \pm 3\%$ compared to $35\pm 4\%$ at $6<z<8$. The overall abundance and decline in the fraction of ``disky'' galaxies between $z=4$ and $z=6$ matches previous studies~\citep{Ferreira2023,Kartaltepe2023}. These studies use an entirely different selection criterion based on visual classification but it is noteworthy that our parametric selection appears superficially consistent.

\begin{figure*}
    \centering
    \includegraphics[width = 0.49\textwidth]{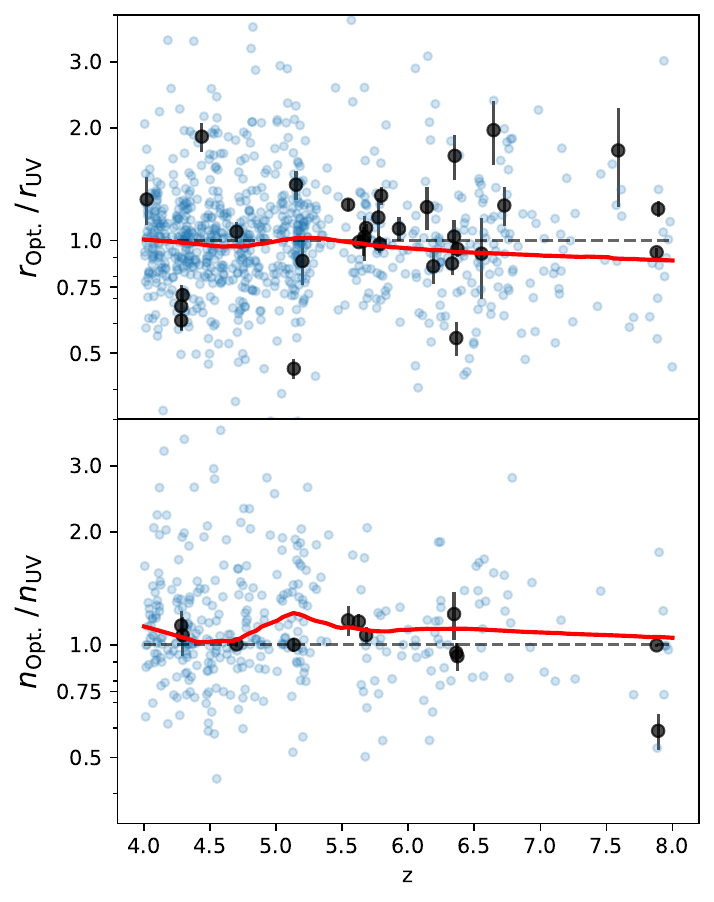}
    \includegraphics[width = 0.49\textwidth]{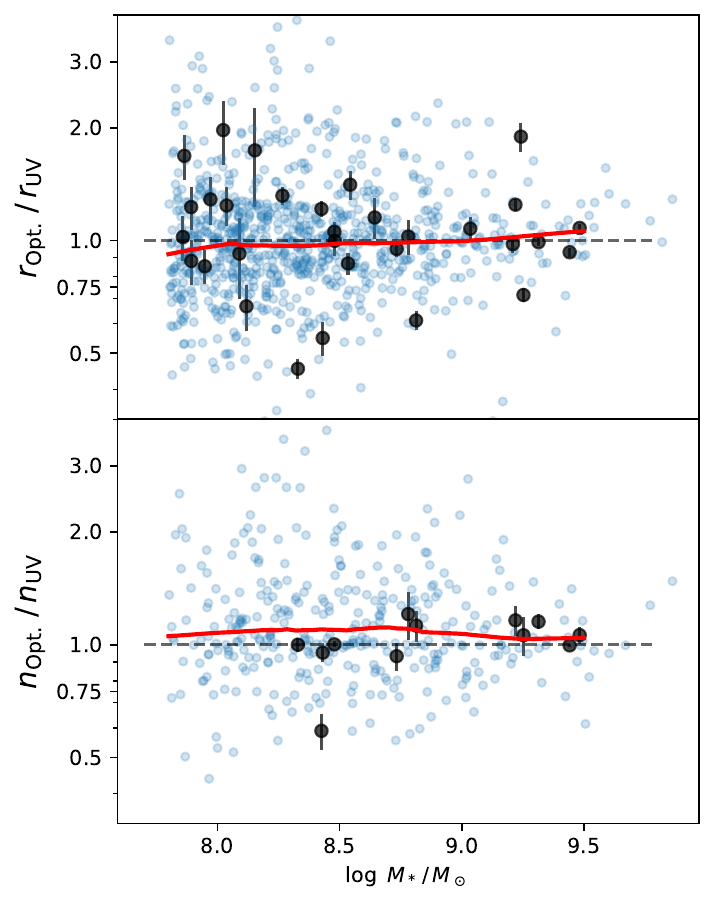}
    \caption{The ratio of effective radius and Sersic index measured at rest-frame optical (5,000 \AA) compared to rest frame UV (2,000 \AA) as a function of redshift and stellar mass. Both are shown on a logarithmic scale with blue points showcasing photometrically selected galaxies and spectroscopically confirmed in black. The \texttt{loess} measured median value, as a function of redshift. We find little difference between the morphology of galaxies in the rest-frame UV compared to the rest-frame optical across the entire redshift and mass range of our sample.}
    \label{fig:uv_opt}
\end{figure*}

To understand the variation in galaxy profiles with redshift, the ratio of $r_{\rm eff}$ and Sersic index measured at restframe optical and UV as a function of stellar mass and redshift are displayed in Figure~\ref{fig:uv_opt}. Similar to above for this analysis we limit the comparison of radii to galaxies with both $SNR_{\rm opt.}$ and $SNR_{\rm UV}$ greater than 5 and the index comparison to galaxies where both SNRs are greater than 15. As with above these cuts may bias the original sample, given the requirement for high UV and optical SNRs.

In each panel of Fig.~\ref{fig:uv_opt} we show the distribution of individual galaxies in our sample, with black points showcasing the spectroscopic sample and blue for the photometric, alongside the \texttt{loess} locally weighted regression. We find the median ratio of radii, $r_{\rm opt.}/r_{\rm UV}$, is consistent with unity across the entire stellar mass and redshift range of our sample. While individual galaxies may deviate significantly from 1, there does not appear to be any systematic trend. Comparing the Sersic index measured at different wavelengths, $n_{\rm opt}/n_{\rm UV}$, we find the longer wavelength optical index to be marginally higher on average, the median ratio is 1.1, without any discernible trend with redshift or stellar mass. This is in contrast to galaxies at cosmic noon~\citep{Rwat2009} and low redshift \citep{kelvin2012} which show Sersic indices that strongly depend on wavelength.

While the median ratio of both these quantities is near unity, there appears to be significant scatter among the population itself, with the optical to UV size ratio reaching as high as 3 or as low as 0.33 for some galaxies. We briefly investigated if the radii ratio correlates with other physical properties measured from SED fitting, such as star formation rate or dust content, but found nothing conclusive that could fully explain the scatter. Nonetheless it may provide interesting clues into the different pathways for galaxy growth at this epoch. In future work we hope to better understand the cause of the wide variation in the radii ratios, specifically using spectroscopy to better understand the physical properties of individual galaxies.

\section{The Galaxy Size-Stellar Mass Relation}
\label{sec:size_mass}
\begin{figure*}
    \centering
    \includegraphics[width = 0.63\textwidth]{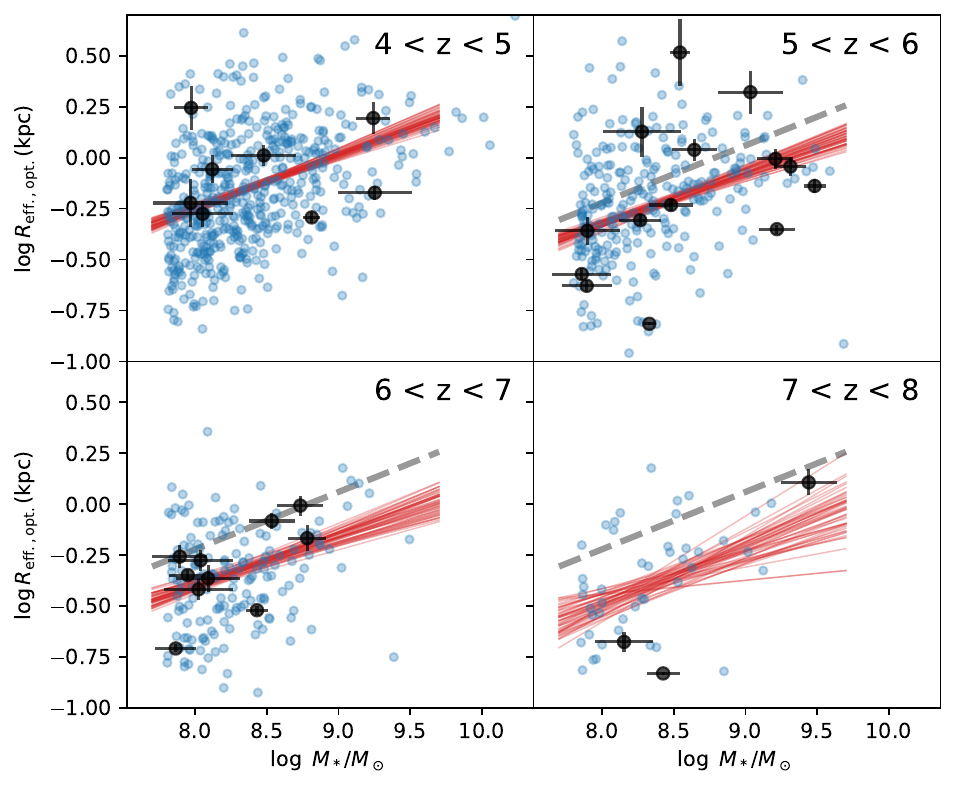}
    \includegraphics[width = 0.33\textwidth]{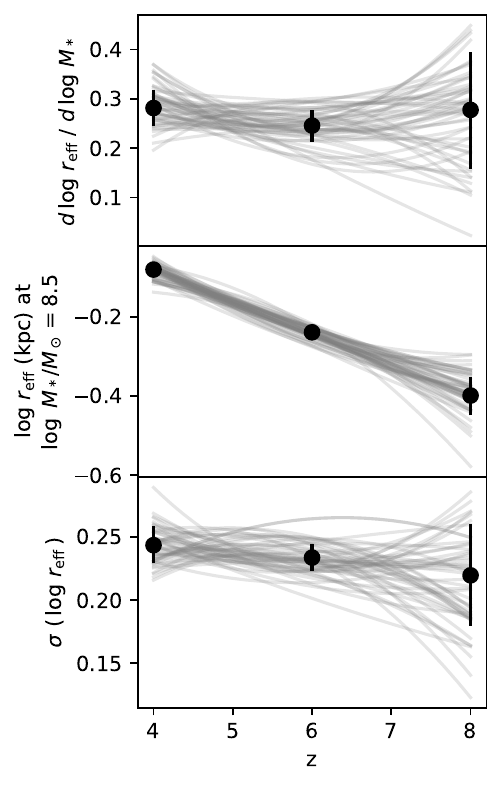}
    \caption{The rest-optical size-mass relation at $4<z<8$. \textit{Left:} The distribution of  optical half-light radius and stellar mass for galaxies in our sample, split into four redshift panels. Photometrically selected galaxies are shown in blue and spectroscopically confirmed galaxies in black. Draws from the posterior distribution of the size-mass fit (See Sec.~\ref{sec:size_mass}) at the median redshift of the bin are plotted in red lines. Grey dashed lines in the three highest redshift panels show the median size-mass at $z=4.5$. \textit{Right:} The derived parameters of the optical size-mass relationship. The knots of the continuity model at $z=4,6, \rm{and}\ 8$ are shown as points with: the median and 16\% - 84\% are plotted. The thin grey lines show 50 draws from the posterior where the quadratic redshift evolution is shown. There intercept is seen to evolve quickly while the slope and scatter stay constant across the redshift range of $4<z<8$.}  
    \label{fig:size_mass_fit}
\end{figure*}

Figure~\ref{fig:size_mass_fit} plots the relationship between stellar mass and half-light radius at optical radii in four discrete redshift bins. The spectroscopic sample is shown in large black points with error bars while the more numerous photometric sample is plotted in small blue dots using the median mass and redshift and median size, corrected for lensing by dividing by $\sqrt{\mu}$. At first glance we see a positive relationship that appears to follow a power law, similar to results at lower redshifts \citep{vanderwel2014, lange2015}. We caution against blindly fitting the observed data points as this approach ignores the plentiful systematic uncertainties. For the photometric sample, uncertainties in the redshifts translate directly to -- likely correlated --uncertainties stellar mass estimates. In addition we have not accounted for uncertainties in the lensing for either sample. In order to accurately infer the characteristics of the size-mass relation, and its evolution with redshift, one must properly account for all of these uncertainties.

\subsection{Fitting Procedure}
\label{sec:size_mass_fit}
To infer the relationship between size and mass we assume it follows a simple power law distribution parameterized by 3 variables: the logarithmic slope, the logarithmic size at $\log M_*/ M_\odot = 8.5$ and a log-normal scatter. While there is growing evidence for a deviation from a power-law for the size-mass relation in certain regimes, this is typically for low-mass quiescent galaxies \citep{nedkova2021, Cutler2024} or massive galaxies $\log M_*/ M_\odot > 10.5 $ \citep{mowla2019b, kawinwanichakij2021}, neither of which are present in our sample. The maximum stellar mass in our sample is $\log M_*/ M_\odot = 10.2$ and $<1\%$ of our lies in the $u_s g_s i_s$ quiescent galaxy selection defined in \citet{antwidanso2022}. Additionally by analyzing the median sizes in Fig.~\ref{fig:size_mass_fit} we see no obvious deviation from a simple power law. 
%We will characterize the size-mass relationship using three parameters: The power law slope, the normalization at \logmsol$=8.5$ and the log-normal scatter.

Instead of simply fitting the size-mass relations independently in different redshifts bins we adopt a continuity model by enforcing the parameters of the size-mass relation to evolve smoothly with redshift~\citep[see][ for e.g.]{leja2020, Morishita2024}. This technique has several benefits:  it enables us to fit all of the galaxies in our sample simultaneously, forces consistent evolution with redshift and naturally allows for the inclusion of photometric redshift uncertainties. The parameters for the size mass relation will be varied at three knots, at $z=4,6, \rm{and}\ 8$. For each parameter we use these knots to define a quadratic function that defines the continuous evolution with redshift. Using knots to define the quadratic function makes it easier to define physically motivated priors for each parameter, rather than e.g. using the coefficients of the quadratic.. The priors for each of the parameters are the same across the three knots; -2 \textendash 1 for the logarithmic size at $\log M_*/ M_\odot = 8.5$ in kpc, -1 \textendash 1 for the slope and $10^{-3} \textendash 1$ for the log-normal scatter.

Within our fitting procedure we account for the uncertainties affecting the measured size, lensing magnifications, stellar mass, and redshift for the photometric sample. When considering the distributions of stellar mass (and photometric redshift) produced by \texttt{Prospector}, we need a way to represent these complex distributions to use in the inference of the size-mass relationship. For this we turn to Gaussian Mixture Models (GMMs), which we fit using the \texttt{mixture} module in \texttt{scikit-learn} \citep{scikit-learn}. For the spectroscopic sample we assume the redshift is fixed to the $z_{spec}$ reported in \citet{Price2024}. The posterior distribution for the stellar mass is modeled as a two-component GMM. For the photometric sample, the posterior surface is often more complicated as the stellar mass and inferred redshift are likely correlated. To fully capture the complex posterior distributions we fit a 4 component Gaussian Mixture Model (GMM) to the joint $z-\log\, M_*$ posterior produced by \texttt{Prospector}. 

For the measured optical size we use the wavelength-dependent half-light radius measured above in Sec.~\ref{sec:sizes}. We assume the posterior for the size is Gaussian and calculate the mean and width at the wavelength corresponding to $5000$\AA\ for a given redshift. For the spectroscopic sample, this is fixed, but for the photometric sample this varies with the uncertainty on the photometric redshift.
For all galaxies we assume the uncertainty on the measured magnification is Gaussian with a fractional uncertainty $10\%$. We ``de-lens'' the measured size of galaxies by decreasing them by a factor of $1/\sqrt{\mu}$. For the photometric sample we account for the change in the lensing and angular diameter distance with the sample redshift, although at these redshift there is not a strong dependence of the lensing with redshift.

For the photometric sample we additionally allow for the possibility of outliers, for examples caused by catastrophic failures in the photometric redshift fit. To marginalize over this possibility we include an additional parameter, the outlier fraction, $q_{\rm outlier}$. Following \citet{hogg2010}, this defines the relative importance of the size-mass fit distribution and an outlier, or background distribution. For the latter we assume any outlier will follow the overall distribution of sizes in the catalog in pixel space.
%To account for this we define an outlier distribution, in pixel space, based on the \texttt{flux\_radius} parameter in the F277W+F356W+F444W detection image. \citep{Weaver2024}. Similar to our original selection we consider all galaxies with $S/N>10$. To (crudely) correct for the PSF we subtract the HWHM size of the F444W PSF in quadrature from the \texttt{flux\_radius}. 
Using \texttt{flux\_radius} measured from the detection image in the \citet{Weaver2024}, catalog we find the overall distribution in pixel space is well fit by a log-normal with a peak at $\log\ R_{\rm eff, pixels} = 0.144$ with a width of 0.333. This will be used as the outlier, or background distribution. Motivated by the $z_{\rm phot}-z_{\rm spec}$ comparison presented in \citet{Suess2024} and \citet{Price2024} we set a prior on the $q_{\rm outlier}$ to be a Gaussian distribution with a mean of 0.07, width of 0.02 and bounded by 0.0 and 0.2. 

The fitting procedure is written in the \texttt{numpyro}. The parameters of the size-mass distribution are sampled using a No U-Turn MCMC sampler using four chains with 1000 warm-up and 2000 sampling steps each. This results in the effective sample size of the posterior distribution for the size-mass parameters are all $> 2000$ and $\hat{r}$ metric, comparing the inter-chain to intra-chain variance, is $<1.02$ suggesting the sampling results are converged and robust.

\begin{table*}
    \centering
    \caption{Inferred parameters for the optical size - stellar mass relationship at the knots of the continuity model.}
    \begin{tabular}{c c c c}
         & $\log\, R_{\rm eff}/\rm{kpc}$ at $\log\, M_*/M_\odot = 8.5$   & $d\log\, R_{\rm eff}/ d\log\, M_*$ & $\sigma(\log\, R_{\rm eff})$ \\
         & (Normalization) & (Slope) & (Scatter) \\ \hline \hline
        $z = 4$ & $-0.08 \pm 0.02$ &$0.28 \pm 0.04$ &$0.24 \pm 0.01$ \\
        $z = 6$ & $-0.24 \pm 0.01$ &$0.25 \pm 0.03$ &$0.23 \pm 0.01$ \\
        $z = 8$ & $-0.40 \pm 0.05$ &$0.28 \pm 0.12$ &$0.22 \pm 0.04$ \\
        \hline
    \end{tabular}
    \label{tab:size_mass_res}
\end{table*}

\subsection{Results}
\label{sec:size_mass_res}

The inferred parameters of the power law fit to the size-mass distribution, combining constraints from both the spectroscopic and photometric sample, are shown in Table~\ref{tab:size_mass_res}. We report the median and $1-\sigma$ intervals for the parameters at the three knots but reiterate that all of the galaxies are fit simultaneously using a quadratic function to model the redshift evolution. This evolution is displayed visually in Fig.~\ref{fig:size_mass_fit}, along with posterior predictive size-mass relations plotted alongside the individual galaxies. As discussed above in Sec.~\ref{sec:size_mass_fit}, we include the possibility of outliers, likely caused by photometric redshift failures and allow the overall fraction to vary. The inferred value of the outlier fraction is $0.03\pm0.02$ in line with previous estimates of the catastrophic failure rate of photometric redshifts for this sample.~\citep{Price2024,Suess2024}

The full joint-posterior distribution of the size-mass parameters are displayed in Appendix~\ref{sec:app_size_mass_res} where we compare different variations of this size mass-fitting procedure. We compare constraints when the spectroscopic or photometric sample are used in isolation and find good agreement. We note that most of the constraining power is coming from the photometric sample, especially at $z>7$ as the number of spectroscopically confirmed galaxies is only three. While the redshifts and stellar masses of the photometric sample are more uncertain, the 20 times larger sample size is key to improving the constraints. We also test not including an outlier model and/or photometric redshift uncertainties. These are both novel additions that may make it more difficult to compare to previous results. We find neither of these, or the combination, affects the results qualitatively. We additionally test for lensing related systematics, such as a bias toward compact sources, by running the inference on a sub-sample of galaxies with $\mu<1.5$ and find consistent results to our fiducial sample.

\begin{figure*}
    \centering
    \includegraphics[width = \textwidth]{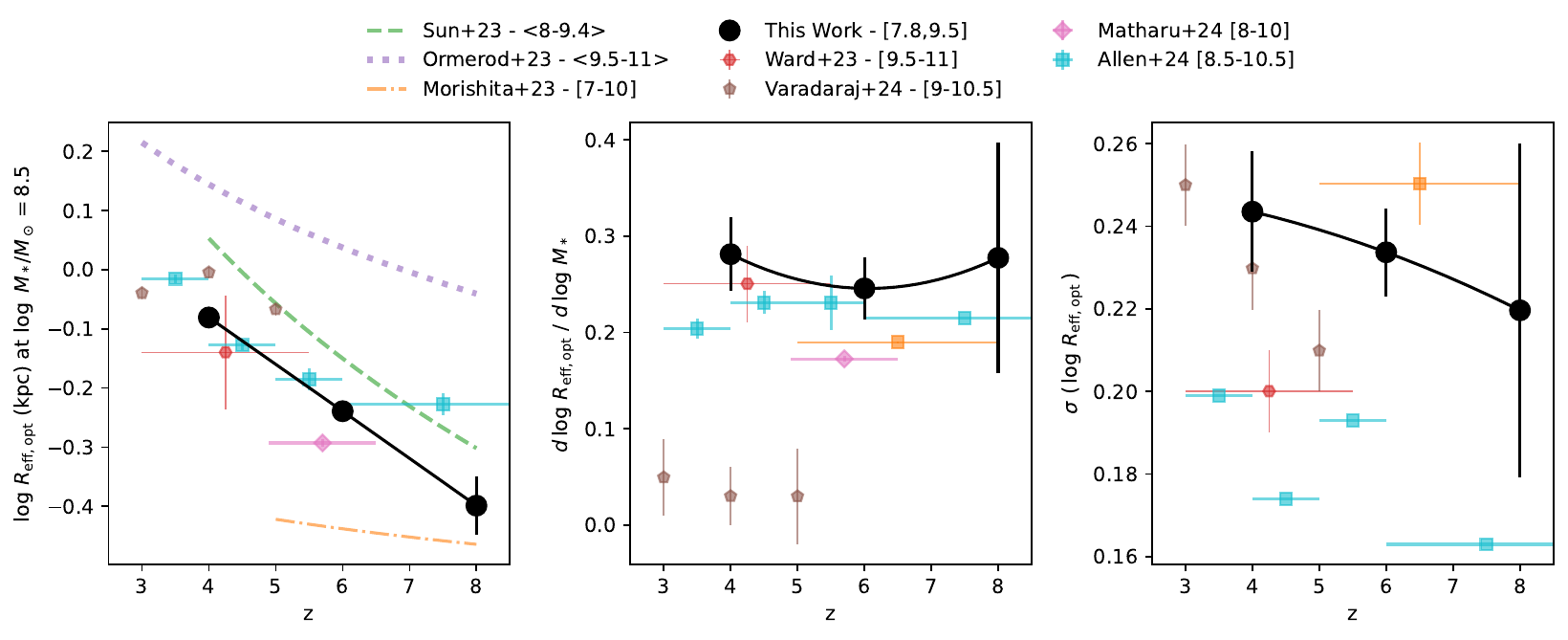}
    \caption{Comparison of the derived parameters of the rest-frame optical size - stellar mass relation at $3<z<8$ using data from JWST. The normalization, slope and logarithmic scatter are shown in separate panels. Our results largely agree with previous studies, which find slope of approximately 0.25 and the logarithmic scatter of around 0.22}
    \label{fig:size_mass_lit_comp}
\end{figure*}

The normalization of the size mass relation is the best constrained parameter in our inference and decreases with increasing redshift. The inferred average size of our sample at \logmsol$=8.5$ decreases from \logrkpc$=-0.08$ at $z=4$ to \logrkpc$=-0.4$ at $z=8$. This implies a factor of two growth in the average size of galaxies in $<1$ Gyr. For perspective, following the parameterization in \citet{vanderwel2014}, the sizes of star-forming galaxies also doubled between z=1.5 and z=0, a period of roughly 9 Gyr.

We find the slope of the size-mass relation to be constant to high precision between $4<z<6$ with an average value of 0.25. This value is similar to many studies at lower redshift which consistently find a positive slope for the size-mass relationship for star-forming galaxies in the range of 0.2-0.3.  At $z>6$, the inferred slope becomes highly uncertain. The median of the posterior at $z=8$, is similar at 0.28, but the $1-\sigma$ uncertainty range is large at 0.12. At $z>7$, there are only a handful of galaxies in our sample with \logmsol$>8.5$, meaning the dynamic range we have to infer the slope is quite small. Combined with the overall decrease in the number of galaxies in our sample at this redshift, the inferred slope is highly uncertain.

We infer the log-normal scatter around the size mass relation to be constant across the entire redshift range of our sample at a value of 0.23. Similar to the inferred slope, the uncertainty on this increases at $z>7$ due to the small number of galaxies. We tested models where the uncertainty was assumed to be a constant with redshift and found a similar inferred value of $0.23\pm 0.01$. The inferred slopes and intercepts were also consistent but we did not find any appreciable benefits in their constraints so have opted to report the flexible model where the scatter is allowed to vary with redshift.

We compare the parameters for the optical size-mass relationship measured in this study to others at $z>3$ that have used JWST data. Given the many differences between all of the studies we aim to provide context by denoting the ranges $\log\, M_*/M_\odot$ used in the legend. If a parametric form was fit to the size-mass relation, as in this paper, it is marked with $<>$ and the average size is extrapolated to \logmsol$=8.5$. $[]$ in the legend denotes studies where some form of the average size is used. Points are used if the study uses discrete redshift bins, with the horizontal error bars denoting the redshift range. Lines are plotted if instead a parametric, continuous form for the redshift evolution is used.

A detailed comparison of our results to \citet{Sun2024} and \citet{ormerod2024} is difficult as these studies only measure an average size, and tend to have higher mass galaxies, therefore larger average sizes, although the trends with redshift are qualitatively similar to what we find. Compared to other studies measuring the parameters of the size-mass relationship we find similar results. The average size at $\log M_*/M_\odot = 8.5$ matches the relations presented in \citet{Varadaraj2024}, \citet{ward2024} and \citet{Allen2024} extrapolated to lower masses. At $z\approx6$ we are consistent with the findings of \citet{Matharu2024} but find a significantly higher normalization, approximately 0.2 dex, compared to \citet{Morishita2024}, although are consistent with the latter study at $z=8$. The slopes at different redshifts match those of \citet{ward2024} and \citet{Matharu2024}\footnote{We note that \citet{Morishita2024} fix the slope of the size-mass to be 0.19 in their analysis. Similarly \citet{Allen2024} fix the slope to be 0.21 for their highest redshift bin at $z>6$}. The results presented here and all of the others report much larger slopes than \citet{Varadaraj2024}, who measure a slope of the size-mass relation near zero at $z\approx4$. The largest obvious difference is that in \citet{Varadaraj2024} the authors select galaxies, and calculate stellar masses, from ground based data whereas our study, and the others shown, select and perform SED fitting on JWST data. However it's unclear if the difference in instrument is enough to cause this discrepancy. We find slightly high scatter, 0.23 dex at $z=6$ compared to \citet{Allen2024} and \citet{ward2024}, by about 0.04 dex but are consistent with \citet{Morishita2024} and \citet{Varadaraj2024}.

%%%% I have a plot comparing r_UV - M_UV for out galaxies to HST measired relations and they match well enough, not sure if it's worth including %%%%
%As an additional comparison we compare out results to inferred $R_{\rm eff, UV}$-$M_{\rm UV}$ relations derived with HST. xxxyyyzzz

\section{Discussion}
\label{sec:disc}
\subsection{The Size Evolution of Low-Mass Galaxies Across Cosmic Time}
\begin{figure*}
    \centering
    \includegraphics[width = \textwidth]{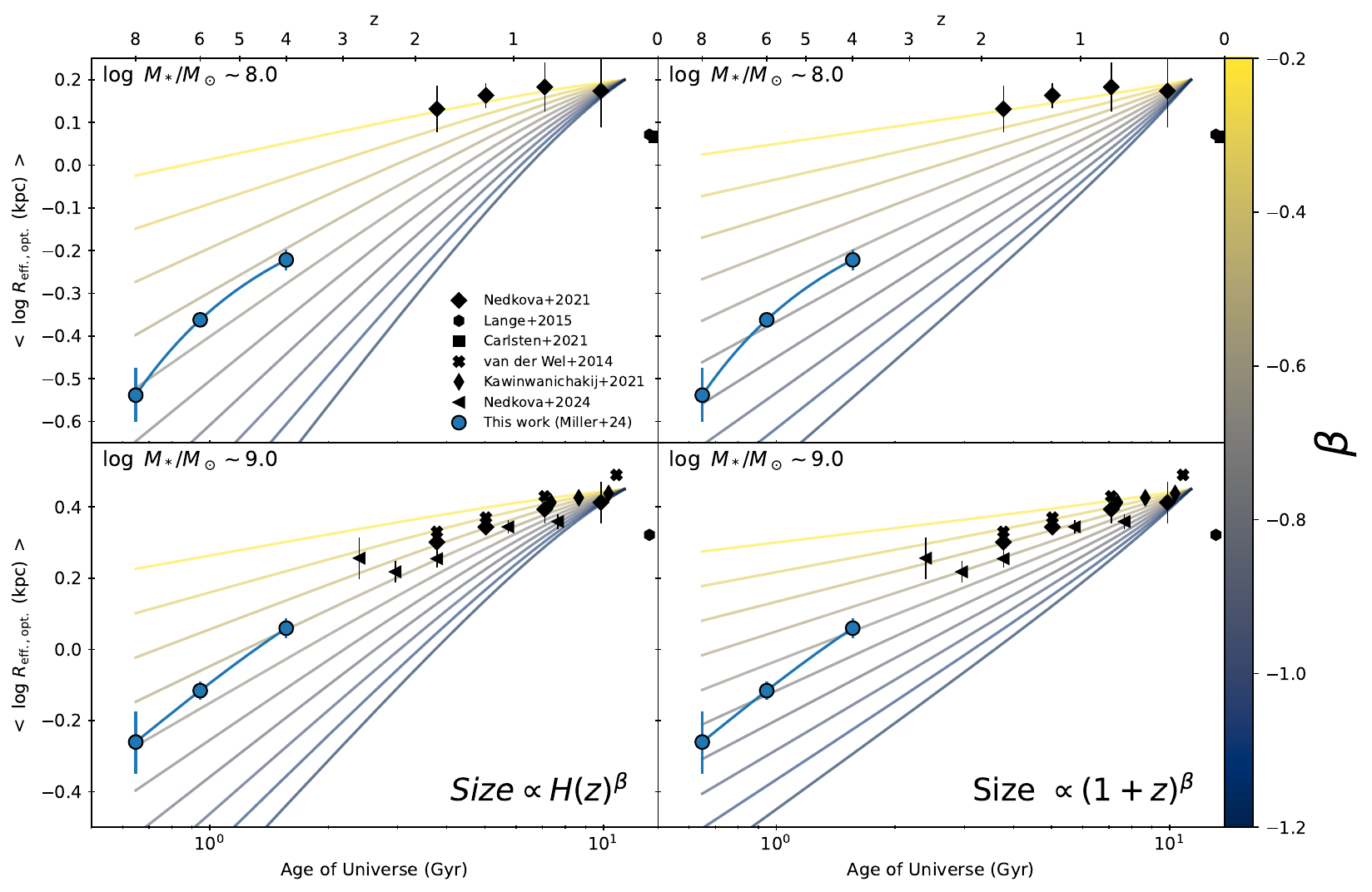}
    \caption{Connecting the sizes of low-mass galaxies across cosmic time. The average optical half-light radius of galaxies at $\log M_* / M_\odot$ = 8 (top) and  $\log M_* / M_\odot$ = 9 (bottom) is plotted as a function of age of the Universe. We compare to previous studies who measure the size-mass relation at $z\lesssim2$ \citep{vanderwel2014,lange2015, Carlsten2021, kawinwanichakij2021, nedkova2021,Nedkova2024}. Common parameterizations for the size-evolution of galaxies, a power law in the Hubble Constant (left) or in $1+z$ (right), with colors representing variations of the power law index. For simplicity all of these curves originate from a fixed point at $z=0.2$. Neither of these parameterizations can explain the small sizes or rapid growth at $z>4$ observed in this study.
    }
    \label{fig:size_evo}
\end{figure*}

The exceptional near-IR imaging capabilities of the NIRCam instrument allows us to study the rest-frame optical morphology of galaxies out to z=8, only 650 Myr after the Big Bang. The combination of joint multi-band modeling and Bayesian methods allows us to robustly determine the morphological properties of galaxies in the early Universe. In Appendix~\ref{sec:recov_tests} we show that sizes can be accurately measured down to a single band SNR of five including reasonably well calibrated uncertainties. Even if the uncertainties on the effective radius of individual galaxies are quite large $\gtrsim20\%$, when used in the size-mass inference we can still robustly determine the characteristics of the population.

In this study we connect the morphology of galaxies from the epoch of reionization to the present using a common metric, the half light radius at optical wavelengths. In Figure~\ref{fig:size_evo} we plot the optical size of galaxies measured in this study compared to previous works which study galaxies at $z\lesssim2$, as a function of age of the Universe. We focus on two masses regimes  for which our data is most sensitive, \logmsol$= 8$ (top panels) and \logmsol$= 9$ (bottom panels). At both masses we again see the rapid growth in  average size in the first 1.5 Gyr of the Universe, growing by about 0.3 dex, or a factor of 2, before $z=4$. By connecting to lower redshift studies we see the rate of size growth decreases, with the average size growing by another 0.4 dex by $z=0.25$, a period of over 9 Gyr. 

We also plot two different commonly used descriptions for the evolution of the galaxy sizes. The left panel displays a model where the average size is proportional to a power law of the evolving Hubble constant, or specific growth rate of the universe~\citep{vanderwel2014}. The right panel shows a model that is proportional to a power law in $(1+z)$~\citep{Shibuya2015}, the inverse of the scale factor. For each parameterization we show the predicted evolution for several different power-law slopes ranging from $-0.2 < \beta < -1.2$, that encapsulates the previously reported values.~\citep{vanderwel2014, Shibuya2015,kawinwanichakij2021, Morishita2024, Allen2024, Varadaraj2024}. 

While these parameterizations explain the size growth of galaxies at $2>z>0.25$ well, neither model with a constant value of $\beta$ can adequately explain the size evolution over all of cosmic time. Especially the rapid growth in the evolution of sizes in the first 1.5 Gyr of cosmic time. For \logmsol$=9$, the compilation presented here at $z<2$ is well described by a power law slope of $\sim 0.6$ for either parameterization. However at $z>4$ this overestimates the findings of our study by 0.2 dex and by 0.4 dex at $z=8$. The parameterizations that match this rapid growth in the early universe are too steep and do not match the measurements at $1<z<2$. For \logmsol$=8$ there is only one constraint at $z>0.25$ by \citet{nedkova2021}, which appears to favor a shallow evolution, with $\beta \sim{}0.2-0.3$. The predictions overestimate the average size of galaxies measured in this study by 0.3 dex at $z=4$.

We also note that at these masses the sizes of galaxies appears to decrease between $z=0.25$ and $z=0$, when compared to studies in the local universe by \citet{lange2015} and \citet{Carlsten2021}. This non-monotonicity is also not captured by either of the discussed parameterizations, which are both monotonic. For this reason we have chosen to normalize these predictions at $z=0.25$ and leave a full exploration of this apparent decrease in sizes at late times to another study.

%% Little bit more about how this is complicated??

The rapid growth of galaxy sizes observed in this study must coincide with a physical mechanism. Two commonly invoked mechanisms to size growth are mergers and inside-out star-formation. The galaxy merger rate is predicted \citep{rodriguezgomez2016} and observed to be higher at this redshift \citep{Duan2024, Rose2024}, however the physics are complicated. Dry minor mergers are generally thought to increase the effective radius through the addition of accreted material to the outskirts~\citep{newman2012}, but in the early universe these mergers will likely be gas rich, which may drive central star-bursts decreasing the size, therefore how these affect $R_{\rm eff, opt}$ is unclear.~\citep{Robertson2006}

The latter, inside out star-formation where more star-formation occurs in the outskirts of the galaxy is supported by the recent findings of \citet{Matharu2024}. These authors study $H\alpha$, a proxy for star-formation rate (SFR), profiles of galaxies out to $4.5< z < 6.5$, using slitless spectroscopy obtained using the FRESCO survey. They find that $H\alpha$ equivalent widths increase at larger radii, suggesting an increase in the SFR of a galaxy in the outskirts. Crucially this excess occurs at all masses probed in their study, as shown in their Figure 3, which is consistent with our findings of self-similar size growth for all masses and a constant slope of the size-mass relation. This inside-out growth is also seen in detailed studies of galaxies at z=7.4 in \citet{Baker2023} and z=5.4 in \citet{Nelson2024}. This is further supported by dynamical modeling results \citep[e.g.][]{degraaff2024b} suggesting that (at least some) galaxies at this epoch are kinematically cold rotating disks which are conducive to this style of inside-out growth.

\subsection{The (lack of) Evolution in the Slope and Scatter of the the Size-Mass Relation}

We find a remarkable consistency in the slope and scatter of the size-mass relation between the early universe and local galaxies. These two properties have often been used to study the connection between galaxies and their dark matter halos. Such as in \citet{shen2003}, where the use the relatively shallow slope, $~0.2-0.25$, of the size-mass relation for star-forming galaxies in the local Universe to argue that the ratio of halo to stellar mass is not a constant, as this would imply a steeper slope of $1/3$, assuming a simple disk model for star-forming galaxies~\citep{Fall1980, mo1998}. \citet{vanderwel2014} measure a similar slope of the slope of the size-mass relation out to $z\sim 2$ which mimics the un-changing stellar to halo mass relation out to this redshift consistent with independent, empirical estimates of the galaxy-halo connection~\citep[e.g.][]{Moster2018, behroozi2019, rodriguezpuebla2017}. 

At $z>4$, the findings from different studies on the stellar to halo mass relation are somewhat contradictory. Some studies find significant evolution \citep{behroozi2013, behroozi2019, Harikane2016}, whereas others find modest or no evolution \citep{rodriguezpuebla2017, Moster2018, stefanon2021}. We note that all of these studies are pre-JWST where constraints on the galaxy-halo connection at this redshift were far more scarce. The consistency of the size-mass slope measured in our study out to $z=6$ may suggest that the stellar to halo mass relation remains consistent out to this redshift. Further work to empirically constrain the galaxy-halo connection in the JWST era are needed to confirm this. At $z>6$, our study provides much weaker constraints on the slope such that we cannot rule out that the slope evolves.

We find that the log-normal scatter stays relatively constant between $4<z<8$, with a measured value of $0.23\pm 0.01$ at $z=6$. This is slightly higher than the consensus at $z<2$, which find scatter of $0.18$ dex~\citep{vanderwel2014, kawinwanichakij2021, Carlsten2021}. This slight increase in scatter could be explained by the expectation that star-formation histories of high-redshift galaxies should be more variable and bursty than at lower redshift~\citep[e.g.][]{Sun2023, Endsley2024}. These bursts cause time variable outflows which can affect the gravitational potential of low-mass galaxies and cause short timescale variations in the galaxy size~\citep{elbadry2017,emami2021}. More work is needed to quantify how much bursty star-formation histories affect the sizes of galaxies and therefore the scatter of the size-mass relation. An additional $\approx 0.15$ dex of scatter is needed to explain this increase, assuming they are added in quadrature. 

The scatter around the size-mass relation is sensitive to systematic and un-accounted for uncertainties. For this study in particular the need to correct for lensing adds additional complexity to our inference procedure. In Appendix~\ref{sec:app_size_mass_res} we test if the lensing corrections imparts any systematic bias on the inferred parameters of the size-mass relation by re-running the measurement procedure on a high and low magnification sample. We find some minor differences between the two samples, at around the $1-\sigma$ significance level, but nothing systematic over the entire redshift range.
%In short, we find no significant difference in the measured scatter (or the slope or intercept) when comparing these samples to our fiducial results. 
This suggests that systematics due to lensing alone cannot account for the difference in scatter measured in this study at $z>4$ and previous works at $z\lesssim 2$. However the methods to infer sizes and mass, and the type of data used varies greatly between our work and those of ~\citet{vanderwel2014, kawinwanichakij2021, Carlsten2021} at lower redshifts. There is potential for unknown systematic differences that could account for the discrepancies in the measured slope or scatter of the size-mass relation. A homogeneous analysis across all redshifts is the next step towards a robust comparison of the morphologies of galaxies across all of cosmic history.

The measured scatter in the size-mass relation is often used as an additional piece of evidence to suggest that sizes of galaxies are related to their halos, specifically their spin or angular momentum~\citep{shen2003,kravtsov2013, vanderwel2014, somerville2018}. In the disk model of formation, it is usually assumed that the variation in formation efficiency is minor such that the size strongly depends only on the halo mass and the spin of the galaxy~\citep{Fall1980,mo1998}. The distribution of sizes at a given stellar mass also appears to match the shape and scale of the distribution of spin parameters, $\lambda_{\rm spin}$, of dark matter halos~\citep{shen2003, somerville2018}. 

The connection between galaxy size and halo spin may be tenuous at high redshift where we expect the galaxy formation process to be chaotic and unordered \citep{Ma2018}. \citet{jiang2018} use two different hydrodyanmical simulations and find that halo spin is not correlated with galaxy size, especially at $z>1$. This is attributed to the diversity of spin and accretion histories, mergers along with compact star-formation phases that decouple the angular momentum of the baryons at accretion onto $R_{\rm vir}$ and the central galaxy itself. \citet{Rohr2022} use the FIRE simulations to study the galaxy sizes - halo size connection and find no evidence for significant dependencies of the galaxy size beyond halo mass, including spin. Specifically at high-redshift, using the \texttt{THESAN} simulations \citet{Shen2024} find that the angular momentum of galaxies is in reasonable agreement with the disk model for simulated galaxies at \logmsol$>9$ but not for lower mass galaxies, which are better described by a spherical shell model. The distribution of galaxy sizes at $z>4$, appears to match the distribution of halo spins but the applicability of the disk model in this chaotic phase of galaxy formation is uncertain. 

Another consideration that we have largely ignored so far is the presence of compact, star-forming galaxies as suggested in \citet{Baggen2023,Baggen2024} and \citet{Morishita2024}. While we have explicitly chosen to remove LRDs~\citep{Labbe2023} and galaxies from our sample that appear as point sources, there are 25 galaxies in our sample with measured sizes $<200$ pc at $z>6$, including two spectroscopically confirmed galaxies. These galaxies are compact, but appear to be resolved in the images, with measured sizes of $\approx 1$ pixel on average with the mean of the posterior distribution of $r_{\rm eff}$ more than 3$\sigma$ away from the lower limit of the prior. These compact galaxies represent about 10$\%$ of the sample, consistent with \citet{Morishita2024}.  Theoretically compact galaxies are thought to be formed by highly efficient central star-formation. Different models suggest this is due to a short lived Feedback free bursts (FFB) \citep{Dekel2023, Li2023}, or because the density is high enough to overcome the momentum imparted by stellar feedback~\citep{Roper2023, Boylan-Kolchin2024}. All of these models are in broad agreement with the observations and predicted sizes of $\mathcal{O}(200 pc)$ at $z=7-8$ and high central star-formation rate densities as seen in~\citet{Morishita2024}. Additionally these models predict the decreased importance of these galaxies at $z<6$. 

If these compact galaxies are formed through a different, highly efficient mechanism it is reasonable to expect that they do not fall on the ``typical'' size-mass relation. Therefore a simple power-law fit of star-forming galaxies could be failing to capture the true complexity of the size-mass distribution. Performing inference using such a parameterization, as done in this study, could lead to biased results such as a lower inferred slope or intercept. Due to the nature of the UNCOVER and MegaScience surveys, we do not have a large enough sample of massive galaxies at $z>7$ in this study to fully explore this question. Current and up-coming photometric surveys that cover a large area such as COSMOS-Web~\citep{Casey2023} and/or spectroscopic surveys that provide a representative sample of massive galaxies at $z>7$ like RUBIES~\citep{deGraaff2024} will allow us to understand the demographics and nature of these compact sources.

\subsection{The Evolution of the Central Density - Mass Relation}

\begin{figure}
    \centering
    \includegraphics[width=\columnwidth]{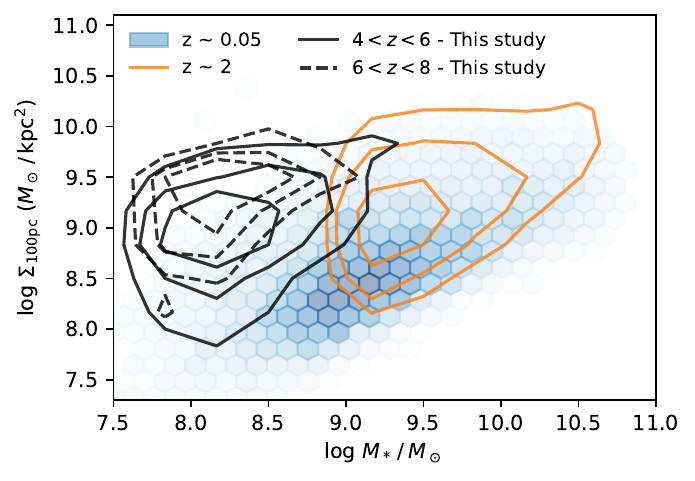}
    \caption{The relationship between central density, measured within 100pc, and stellar mass. Grey contours show the distribution of galaxies at $z>4$ in this study split into two redshift bins. We compare from star-forming galaxies  our sample at $z>4$, at $z\sim 2$~\citep{suess2021} and at $z\sim0.05$~\citep{kelvin2012, lange2015}. Galaxies at $z>4$ show a positive relationship with a similar slope to later epochs with an evolving intercept.}
    \label{fig:density_comp}
\end{figure}

\citet{suess2021} present another physical interpretation of the mass-size relation, suggesting it is a projection of the central density to mass relation. Using 7,000 galaxies at $1<z<2.5$, the authors separate galaxies into evolutionary stages based on their spectral shapes. Each group scatters similarly around the density-mass relation but carve out steep relations in the size-mass plane and have larger scatter. \citet{suess2021} argue that this implies the density-mass relation is more fundamental than the size-mass relation.

Figure~\ref{fig:density_comp} examines this theory in the context of the central-density mass relation for our sample. We also show the star-forming galaxies in the cosmic noon sample used in \citet{suess2021} as well as a sample of star-forming galaxies in the local universe, derived using the GAMA catalog~\citep{kelvin2012, lange2015}.
%derived from ATLAS 3-D \citep{Cappellari2011,cappellari2013} and GAMA \citep{kelvin2012, lange2015} respectively.
We note that although it is common to use $\Sigma_{\rm 1 kpc}$, the density with 1 kpc, since many of our galaxies have $r_{\rm eff} << 1$ kpc, we opt for $\Sigma_{\rm 100 pc}$ instead as it should be more physically meaningful in this context. In the calculation of $\Sigma_{\rm 100 pc}$ we assume $n=1$ however we note that the exact choice of the profile shape does not qualitatively change the conclusions, see the appendix of \citet{suess2021} for a further discussion. For our sample we assume that the optical light traces the stellar mass distribution, which we argue is reasonable as there is no observed color gradient between the UV and optical observed.

We find a positive relation between central density, $\Sigma_{\rm 100 pc}$, and stellar mass for our sample as seen in \citet{suess2021} and the $z=0.05$ sample from GAMA plotted here. %Local elliptical galaxies, which are offset above the relation to high densities as is seen with quiescent galaxies at cosmic noon~\citep{suess2021}.
Qualitatively We find a similar slope of the density for the galaxies at $4<z<8$ compared to lower redshifts, across the limited mass range of our sample. 

To quantify the intercept of each sample shown in Figure~\ref{fig:density_comp} we fit a power law to each sample while fixing the slope to 0.88, as found in \citet{suess2021}. To estimate the uncertainty we perform bootstrap resampling for 1,000 iterations and report the mean and standard deviation. For the low redshift GAMA sample we find, at $\log M_*/M_\odot = 9$, $\log \Sigma_{\rm 100 pc} = 8.51 \pm 0.01$ and for the \citet{suess2021} sample at $z\approx2$, we measure $\log \Sigma_{\rm 100 pc} = 8.8 \pm 0.01$. We split the galaxies in our sample into two redshift bins, at $4<z<6$ we measure $\log \Sigma_{\rm 100 pc}$ at $\log M_*/M_\odot = 9$ to be $9.57\pm 0.02$. For galaxies at $6<z<8$ the central density continues to increase to $\log \Sigma_{\rm 100 pc} = 9.90 \pm 0.03$.

At a given mass, we find galaxies at $z>4$ higher central densities at a given mass, 0.7 dex compared to the $z\sim2$ sample and 1 dex compared to the $z=0.05$ sample. This is contrasted to the only 0.3 dex difference between $z\sim2$ and $z\sim0.05$. By separating our sample into two redshift bins, above and below $z=6$, we find a 0.3 dex increase in the central density for galaxies at $z>6$.

At cosmic noon, the observed tight relation with mild redshift evolution between $\Sigma$ and $M_*$, and the relatively mild size evolution has lead some to suggest that galaxies grow along the $\Sigma - M_*$ relation~\citep{Tacchella2016,Barro2017,nelson2019, suess2021}. We find significant  evolution in the intercept of the $\Sigma_{\rm 100 pc}$- $\log M_*$ that reflects the evolution in the intercept of the size-mass relation. Again this is suggestive of a different growth mechanism compared to cosmic-noon. 

%Even so we find the slope of the density-mass relation does not significantly change out to $z=8$.
%Central density is also seen to correlate strongly with quiescence~\citep[e.g.]{vandokkum2015 ,whitaker2017}

\subsection{Comparison to Theoretical Predictions}
\begin{figure*}
    \centering
    \includegraphics[width=\textwidth]{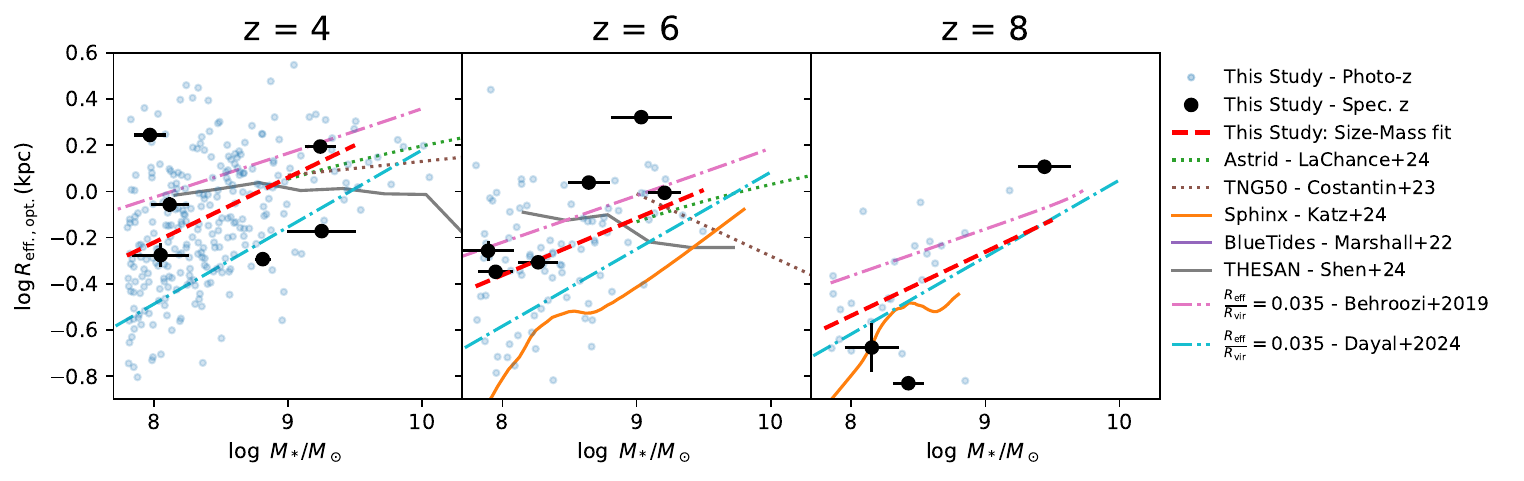}
    \caption{The size-mass relationship measured in this study compared to predictions from hydrodynamical simulations. Individual galaxies from this study are shown in light blue (photometric) or black (spectroscopic). The median parameters of the size-mass relationship are shown in each redshift panel with a red dashed line. We plot parameterized predictions for the size-mass relationship from the Astrid simulations~\citep{Lachance2024} and TNG50~\citep{Costantin2023} in dotted lines. These two studies provide the most apples-to-apples comparison as they forward model images and perform Sersic profile fitting. Predictions for the median galaxy size at a given mass from the Sphinx~\citep{Katz2023} and THESAN~\citep{Shen2024} are shown in solid lines. These two studies on the other hand report the radius of a circle containing half the light, without accounting for observational effects making direct comparison challenging. We see reasonable agreement between observations and simulations at $z<6$ and \logmsol$\approx9$ where the stellar-mass ranges overlap. We also show predictions from two models of the galaxy-halo connection~\citep{behroozi2019,Dayal2024} assuming a constant factor converting from virial radius to effective radius. While the exact conversion factor is uncertain we find good agreement in the evolution and slope of the inferred size-mass relation, which are not sensitive to this choice. }
    \label{fig:sim_comp}
\end{figure*}

In the lead-up to JWST there has been a renewed interest to simulate and predict the evolution of galaxies in the first billion years of the Universe. Many of these studies then forward model observations in order to make predictions for the morphology of galaxies at $z>4$. The observational results in this study are compared to theoretical predictions of the optical half-light radii in Figure~\ref{fig:sim_comp}. For our study we show the distribution of galaxies within $\Delta z = 0.25$ of $z=4,6$ and $8$ along with the median size-mass fit at each redshift.

We also plot published predictions for the parameterized relationship between galaxy size and mass from \citet{Costantin2023} and \citet{Lachance2024}. These studies forward model images, adding realistic observational effects and perform Sersic profile fitting. These represent the most apt comparison to our observation. For \citet{Shen2024} and \citet{Katz2023} predictions, which appear as solid lines, we plot the median galaxy size as a function of stellar mass. Importantly these two studies measure the circularized half-light radii directly from``perfect'' mock images without accounting for observational effects such as noise or the PSF. These differences in methodology can cause large differences in interpretations \citep[see e.g.][]{degraaff2022} so we caution against naive, direct comparisons.

In general we find reasonable agreement with simulations where there is overlap in galaxy samples at $z<6$ and \logmsol$\approx8.5-9$. The larger volume simulations like Astrid and TNG50 have larger mass resolutions and only report sizes down to \logmsol$\approx 9$. Unfortunately this means there is little overlap with our sample that is concentrated at \logmsol$< 8.5$. However, there is good agreement both at $z=4$ and $z=6$ at $\log M_*/M_\odot = 9$ but we do not have enough massive galaxies in our sample to test the flattening slope at higher masses. Recently \citet{Allen2024} reported a slope of the size-mass relation consistent with ours, see Fig.~\ref{fig:size_mass_lit_comp}, extending out to \logmsol$\approx10.5$, seemingly in contradiction with the simulation predictions. The same downturn at high masses is seen in \citet{Shen2024} using the THESAN simulations, which has a finer mass resolution, and broadly agrees with our observations down to $\log M_*/M_\odot = 8.5$. The slope of the size-mass relation from the SPHINX simulations, presented in \citet{Katz2023} shows a similar slope to what is observed in this study at $z>6$ but is systematically offset to lower radii. However due to the difference in methods of how radii are measured the comparison is non-trivial.

Given the large diversity of numerical implementations and included physics, all of the simulation studies make a number of fairly consistent predictions. One of the most striking is that the half-mass sizes - stellar mass relations at $z\gtrsim6$ should be flat and/or negatively slope, specifically at \logmsol$\gtrsim9$.~\citep{marshall2022, Shen2024} This is in stark contrast to lower redshift studies, which even when measuring half-mass radii, always measure a positive slope~\citep{suess2019,mosleh2020, miller2022, vanderwel2024}. Using the Thesan simulations \citet{Shen2024} suggest that these compact sizes for massive galaxies are driven by disk instabilities which lead to highly efficient star formation in the center. Similarly in studying the FLARES simulation, \citep{Roper2022} find the negative slope to be due to compact massive galaxies at $z>5$ created by highly efficient star-formation in the core driven by high local enrichment, leading to efficient cooling. As discussed above, we also find a number of galaxies at $z>6$ which have very compact sizes of $R_{\rm eff} < 200$ pc.

In contrast, these simulations predict the optical size-mass relation that is observed should have a positive slope. This difference between the intrinsic and observed sizes is thought to arise  largely because of differential dust attenuation. These studies forward model the results of the simulations using radiative transfer techniques to measure sizes as they would be observed. The intrinsically brightest regions of the simulated galaxies are also the most affected by dust extinction, leading to more spatially extend light that is less centrally concentrated than the mass and therefore larger half-light radii.~\citep{marshall2022,Shen2024} This is especially important when measuring UV sizes where dust extinction is more important. This leads to the predictions that the slope of the size-mass (or size-luminosity) relation should depend very strongly on wavelengths and become more positive at lower wavelengths.~\citep{marshall2022, Roper2023}. Observationally this is consistent with measurements far-infrared sizes (which trace the cold dust) being smaller on average than the rest-UV or rest-optical sizes, out to at least $z=6$.\citep{fujimoto2017}. \citet{Wu2020}, using the SIMBA simulation, find similar differential dust attenuation but that it is by a negative age gradient leading to similar morphology between UV and optical wavelengths. Additionally in the Astrid simulations, \citet{Lachance2024}~find that optical half-light radii is a good proxy for half-mass radii, except for galaxies with masses \logmsol$>10$.

However, observationally we find no systematic difference when comparing the UV and optical morphology. When comparing the results for individual galaxies, as shown in Figure~\ref{fig:uv_opt}, the average ratio of $r_{\rm opt}/r_{\rm UV}$ is consistent with unity and there is only a modest difference, in measured Sersic index, with the optical index being 10\% high on average. When comparing fits to the size-mass relation we find they agree to high precision out to $z=6$, $0.25\pm0.04$ and $0.22\pm0.04$ for optical and UV wavelengths respectively. At high redshifts, the inferred slope in our study is very uncertain for both wavelengths. Our findings agree with other studies of galaxy morphology at $z>4$ which find little to no difference between optical and UV morphology.~\citep{Ono2023, Treu2023, Morishita2024, Allen2024, Ono2024}. 

At lower redshifts ($z\lesssim2$) there are differences between optical and UV morphology ~\citep{kelvin2012, wuyts2012, Cheng2024}, often found to be caused by radially varying dust attenuation \citep{wang2017,miller2022}. However, this effect is stellar mass dependent. \citet{Nedkova2024} study the ratio of optical to UV radii at $0.5 < <  3.0$ using data from the UVCANDELS survey. This study finds a systematic increase in $R_{\rm UV}\ /\ R_{\rm opt.}$ at high masses , $M_* / M_\odot >10$. By forward modeling simulated galaxies with and without the effects of dust, the authors conclude that the results can be explained solely by the presence of dust and is consistent with other work suggesting more massive galaxies have more dust and obscured star formation~\citep{Whitaker2017b}. For low mass galaxies ($\log M_* / M_\odot \lesssim 10$) they find no significant difference between $R_{\rm UV}$ and $R_{\rm opt.}$, consistent with the results in this study. Combining our results with those of ~\citet{Nedkova2024} suggests that for low mass galaxies across all epochs the average ratio of $R_{\rm UV}\ /\ R_{\rm opt.}$ is near unity, notably with significant scatter.

%This similarity is not present at lower redshifts, $z\lesssim2$, where galaxies exhibit large differences between their UV and optical morphologies~\citep{kelvin2012, wuyts2012}, often found to be caused by radially varying dust attenuation \citep{wang2017,miller2022}. At early times it appears that this radially varying structure has not yet formed.

The root cause of this disagreement between simulations and observations is unclear. It could be a result of missing or imperfect physics that are put into the simulation. The forward modeling process used to assign observational quantities to the simulations is also highly uncertain. Specifically the details of the dust, such as the attenuation curve, grain size distribution, its relative abundance and spatial distribution compared to the gas is extremely impactful and yet highly uncertain. Given a lack of constraints, most of these studies add dust in a post-hoc manner by assigning dust that follows the gas distribution using a simple dust-to-metal ratio. The formation and destruction pathways for dust in the early universe are still highly uncertain and could vary greatly from the prescriptions assumed, many of which are based on results in the Milky Way. Given the sensitivity of UV light to dust attenuation this could greatly affect the predicted morphology. If the dust is treated accurately it could also arise from an offsetting gradient in stellar populations as suggested by~\citet{Wu2020}. Observations~\citep{Baker2023,Nelson2024,Matharu2024} suggest radially varying stellar populations exist at these redshifts however this effect is seemingly only present in the SIMBA simulation used by \citet{Wu2020}.

Given the expected relation between galaxy and halo radii we also plot predictions from two models of the Stellar Mass - Halo-mass relation. First from \citet{Dayal2024} we use Model A for CDM halos, which uses an analytical calculation of the halo mass function and a star-formation efficiency that is tuned to match the high mass end of the stellar mass function at $6<z<10$. We also use the constraints for the \texttt{UNIVERSEMACHINE}~\citep{behroozi2019} an empirical model for the galaxy-halo connection that is fit to a plethora of observations across a range of redshifts. Specifically we use the tabulated values for the stellar mass halo-mass relation for central star-forming galaxies. We then convert halo mass to virial radius using the critical density of the Universe and the evolving definition of the spherical overdensity from ~\citet{Bryan1998}. For simplicity we assume a constant conversion between virial and effective radius at a factor of 0.035. This is supported by empirical and theoretical work with the exact factor ranging from 1.5\% to 5\%~\citep{shen2003, kravtsov2013, somerville2018, mowla2019b,Rohr2022}.

We find that both of the models of the galaxy-halo connection match reasonably well with the observations. One should not read too much into the exact normalization, as we have simply chosen a number within the range expected from previous studies. However the evolution with redshift and the slope of both models qualitatively agree with these predictions and are insensitive to the exact choice of conversion factor. This agreement is further bolstered by results from Hydrodynamical simulations, like FIRE~\citep{Ma2018b,Feldmann2024}, which predict a stellar to halo mass relation with a power law slope of $\sim 1.3 - 1.5$ which is non-evolving at these epochs. When convert this to an virial radius estimate and assume a constant scaling to the effective radius we obtain a size-mass slope of 0.2-0.25, in good agreement with the observations. We also consider the gas radius, $r_{\rm gas} \sim 4.5 \lambda r_{\rm vir} \sim 0.18 r_{\rm vir}$ (assuming $\lambda = 0.04$), as an effective upper limit for the stellar distribution permitted. We find almost all of our galaxies lie beneath the limit for both models. The fact that we find this level of agreement without much tuning is further indication of an inherent connection between a galaxy radii and their halo.

\section{Conclusions}
\label{sec:conc}
In this work we study the sizes and morphology of galaxies in the first billion years of the Universe in the field of Abell 2744. Using data from the UNCOVER and MegaScience JWST programs, we measure half-light radii and Sersic indices for 995 photometrically selected and 35 spectroscopically selected galaxies at $4<z<8$. These measurements using \texttt{pysersic} are made by jointly modeling NIRCam images from six broadband filters using a low-order spline to connect the morphological quantities across the entire wavelength range. MCMC sampling is used for all galaxies to properly assess uncertainties on the measured morphological parameters. We show that this results in 50\% lower uncertainties compared to modeling individual bands individually.

Galaxies exhibit a wide range of Sersic indices at optical wavelengths, with no clear trend with stellar mass. We do find the fraction of galaxies with exponential disk-like profiles, $n_{\rm opt} < 1.5$, decreases from $z=4$ to $z=8$, broadly consistent with results from visually classified galaxies~\citep{Ferreira2022, Kartaltepe2023}. Similar to previous studies using JWST we find no systematic difference in the morphology of galaxies between the rest frame optical and UV. The ratio of radii is consistent with unity across all masses and redshifts ,while the optical index is only 10\% higher on average. Unlike the observations, some studies using hydrodynamical simulations predict that at this redshift the morphology should depend strongly on wavelength~\citep{marshall2022, Roper2023, Shen2024}.

To measure the rest-optical size-mass relation we simultaneously fit a power-law model to all of the galaxies in our sample using a continuity model to allow the parameters to vary smoothly with redshift. Our procedure accounts for uncertainties in the photo-z, stellar mass, lensing and measurement uncertainties on the size itself. We find the normalization of the size-mass relation evolves quickly over the redshift range probed; the average size of galaxies at \logmsol$=8.5$ increases by a factor of 2 in less than 1 Gyr. Our results suggest the logarithmic slope of the size-mass relation is constant at a value of $0.25\pm 0.04$ out to $z=6$. Due to the lack of galaxies in our sample at $z>7$, especially at \logmsol$\gtrsim9$ the measured slope at the highest redshifts probed is highly uncertain, $0.28\pm0.12$. We infer the log normal to be constant across the redshifts range of our sample at a value of $0.23$ dex.

By connecting our results to previous studies at lower redshift, we study the size evolution of galaxies at \logmsol$=8$ and \logmsol$=9$ over 13 billion years of cosmic history. Common descriptions for the evolution of galaxy sizes, such as power laws in $H(z)$, the specific growth rate, of $(1+z)$, the inverse scale factor fail to match the relatively small sizes measured in the study and steep decline at $4<z<8$, when they are calibrated to results at $z<2$. This discrepancy suggests an alternative growth mechanism for low mass galaxies at these early times.

The slope and scatter of the size-mass relation appear to be remarkably consistent over the history of the Universe. The measured slope at $z=6$ of $0.25\pm 0.04$ is consistent within uncertainties to those at $z=0$~\citep{lange2015, Carlsten2021}. The scatter is slightly higher, at a value of $0.23\pm 0.01$ dex, compared to values in the range of $0.14-0.18$ dex \citep{vanderwel2014, Carlsten2021, kawinwanichakij2021}. We discuss the possible explanations for this consistency in the context of the connection between galaxy sizes and their dark matter halos.

We have provided a census of galaxy morphology for low-mass galaxies, \logmsol$<9.5$ at $4<z<8$ that will be complimented by future studies to understand the growth of the first generation of galaxies. We have shown that the average size of galaxies evolves quickly at $4<z<8$ compared to lower redshifts, while the slope and scatter appear relatively consistent across the history of the Universe. With this holistic view we can begin to understand the physical processes that shape low-mass galaxies throughout their entire life cycle.

\begin{acknowledgments}
TBM was supported by a CIERA Postdoctoral Fellowship. This work used computing resources provided by Northwestern University and the Center for Interdisciplinary Exploration and Research in Astrophysics (CIERA). This research was supported in part through the computational resources and staff contributions provided for the Quest high performance computing facility at Northwestern University which is jointly supported by the Office of the Provost, the Office for Research, and Northwestern University Information Technology. PD acknowledges support from the NWO grant 016.VIDI.189.162 (``ODIN") and warmly thanks the European Commission's and University of Groningen's CO-FUND Rosalind Franklin program. AZ acknowledges support by Grant No. 2020750 from the United States-Israel Binational Science Foundation (BSF) and Grant No. 2109066 from the United States National Science Foundation (NSF); and by the Israel Science Foundation Grant No. 864/23. We thank the anonymous referee for their constructive feedback which improved the manuscript.
\end{acknowledgments}

\vspace{5mm}
\facilities{JWST~\citep{Rigby2023} (NIRCAM~\citep{Rieke2023}, NIRSpec~\citep{Boker2023})}

%% Similar to \facility{}, there is the optional \software command to allow 
%% authors a place to specify which programs were used during the creation of 
%% the manuscript. Authors should list each code and include either a
%% citation or url to the code inside ()s when available.

\software{astropy \citep{astropy2018}, 
ArviZ \citep{Kumar2019},
corner \citep{corner},
jax \citep{jax2018},
interpax \citep{interpax},
matplotlib \citep{matplotlib},
numpyro \citep{phan2019},
numpy \citep{numpy},
pandas \citep{pandas},
photutils \citep{photutils},
scikit-learn \citep{scikit-learn},
scipy \citep{scipy},
Source Extractor \citep{bertin1996,sep},}

%% Appendix material should be preceded with a single \appendix command.
%% There should be a \section command for each appendix. Mark appendix
%% subsections with the same markup you use in the main body of the paper
%% Each Appendix (indicated with \section) will be lettered A, B, C, etc.
%% The equation counter will reset when it encounters the \appendix
%% command and will number appendix equations (A1), (A2), etc. The
%% Figure and Table counter will not reset.

\appendix

\section{Parameter Recovery Tests}
\begin{figure*}
    \centering
    \includegraphics[width=0.47\textwidth]{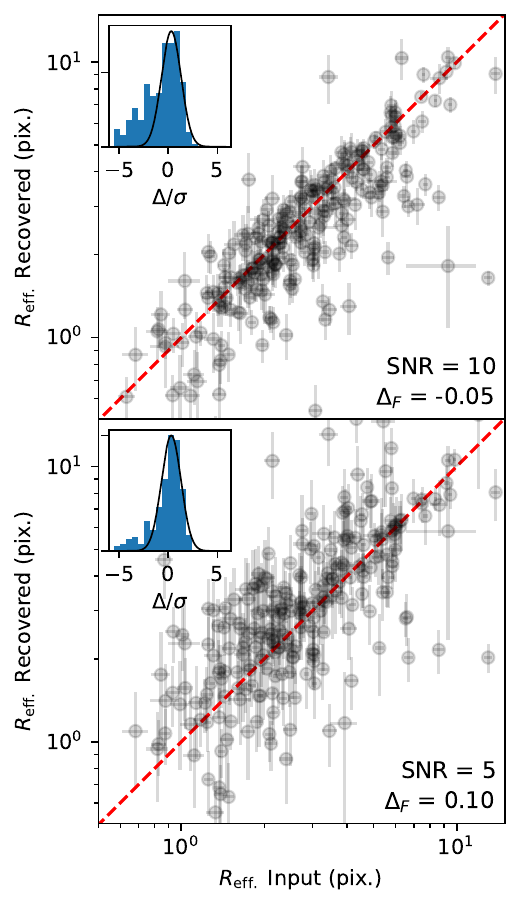}
    \includegraphics[width=0.47\textwidth]{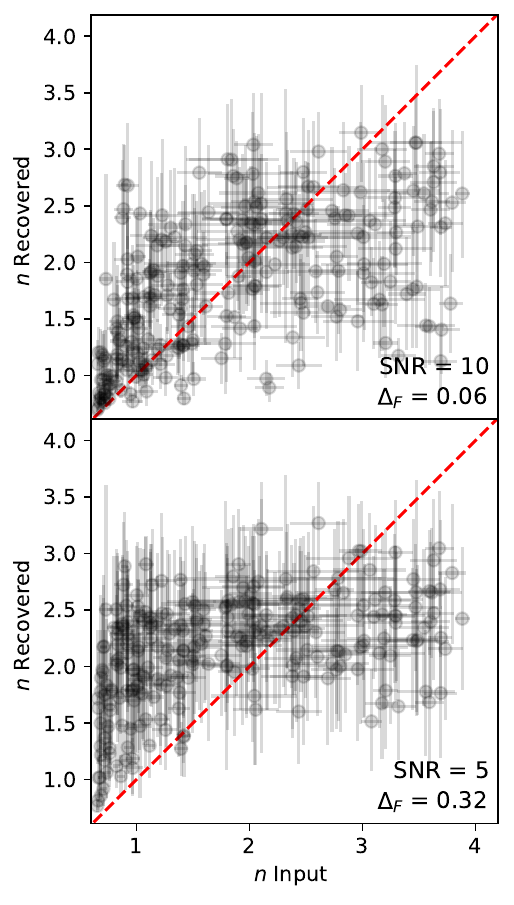}
    \caption{Measured values of the effective radius ($R_{\rm eff}$, left) and Sersic index ($n$, right) recovered from degraded images are compared to those measured from the original image. There are two different samples for which noise is added to achieve a target SNR of 5 (bottom) and 10 (top). In each panel the dotted red line showcases the one-to-one relation and the median fractional difference ($\Delta_F$) is shown. An inset is included in the $R_{\rm eff}$ panels showcasing the distributions of residuals normalized by their respective measurement uncertainties. We find $R_{\rm eff.}$ is able to be successfully recovered down to SNR$=5$, however at these low SNR values no meaningful constraints can be placed on the Sersic index of galaxies. }
    \label{fig:recov_tests}
\end{figure*}
\label{sec:recov_tests}
In this section we perform parameter recovery tests to ensure that our method is able to recover morphological parameters even at low SNR. We begin with a set of 279 high SNR galaxies ($30 < SNR_{\rm opt.} < 50$). For each galaxy we then add add uncorrelated gaussian noise to each image to increase the median background RMS by a factor of $SNR_{\rm opt.}\, /\, SNR_{\rm target}$ to achieve target SNRs of 10 and 5. The same factor is applied to all six filters used in the multi-band fitting. These degraded images are then re-fit using the exact same procedure, except using the masks and initial guesses derived from the original images. In our comparisons we focus on the filter that is closest to rest-frame optical. In Figure~\ref{fig:recov_tests} we analyze how well the morphological parameters $R_{\rm eff}$ and $n$ are recovered.

Focusing on $R_{\rm eff}$, we find consistent recovery even at target SNR of 5. The galaxies in our sample scatter around the one-to-one line and the median fractional difference between the input and recovered values is $-5\%$ and $10\%$ for target SNRs of 10 and 5 respectively. The distribution of residuals normalized by the measured uncertainty, where the input and recovered uncertainties are added in quadrature, showcase largely a Gaussian distribution except for at values of $<-2.5\sigma$. In our sample there is an overabundance of galaxies for which the input $R_{\rm eff.}$ is high and the recovered $R_{\rm eff.}$ is low. This deviation is more apparent in the SNR$_{\rm target} = 10$ sample. Looking in detail at some of the largest outliers we find that there is usually a secondary component or additional flux at large radii, that is present in the original images but covered by the noise in the degraded images. This model mismatch of clumpy galaxies compared to a smooth Sersic profile is likely causing the uncertainties to be somewhat underestimated for these sources. Still, our pipeline is able to robustly recover sizes and their uncertainties for most galaxies in our sample down to SNR = 5.

The comparison on Sersic indices $n$ showcases less consistent recovery. The indices are not able to be constrained from the degraded images and the posteriors are largely wide and cover the entire prior range for 0.65-4. For these low SNR sources the measured values of $n$, especially when the uncertainties are not considered, do not relay any meaningful information. 

\section{Variations of the Size-Mass Fitting Procedure}
\label{sec:app_size_mass_res}
\begin{figure*}
    \centering
    \includegraphics[width = \textwidth]{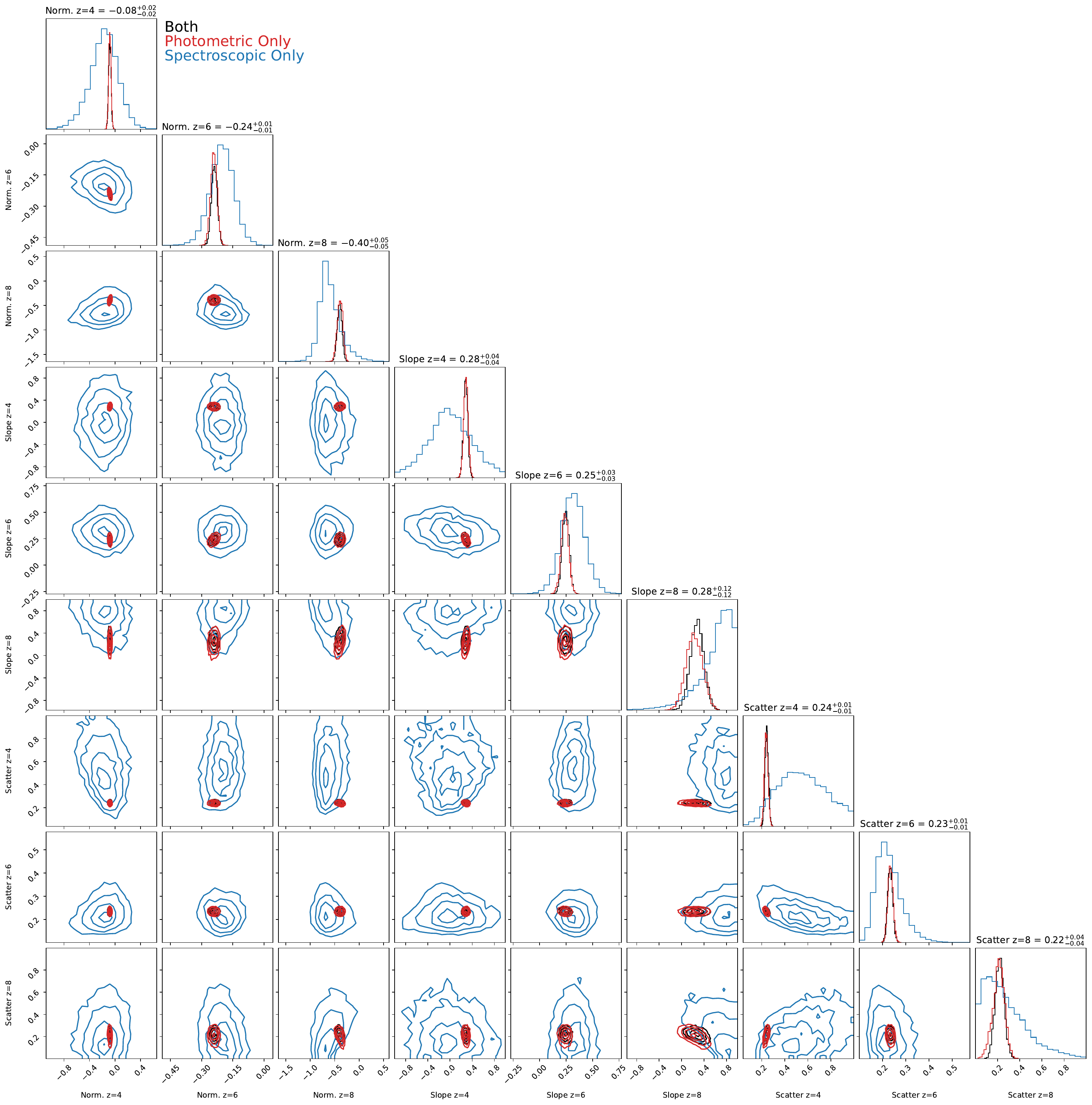}
    \caption{The posterior distribution for the inferred parameters of the size-mass relationship. The joint distributions of each parameter (Normalization, slope and scatter) at each knot of the continuity model is shown in a corner plot. The contours correspond to the 0.5,1,1.5 and 2 $\sigma$ regions of each distribution.  We compare the posterior using the photometric sample, spectroscopic sample and both combined.}
    \label{fig:spec_phot_post_corner}
\end{figure*}

\begin{figure*}
    \centering
    \includegraphics[width = \textwidth]{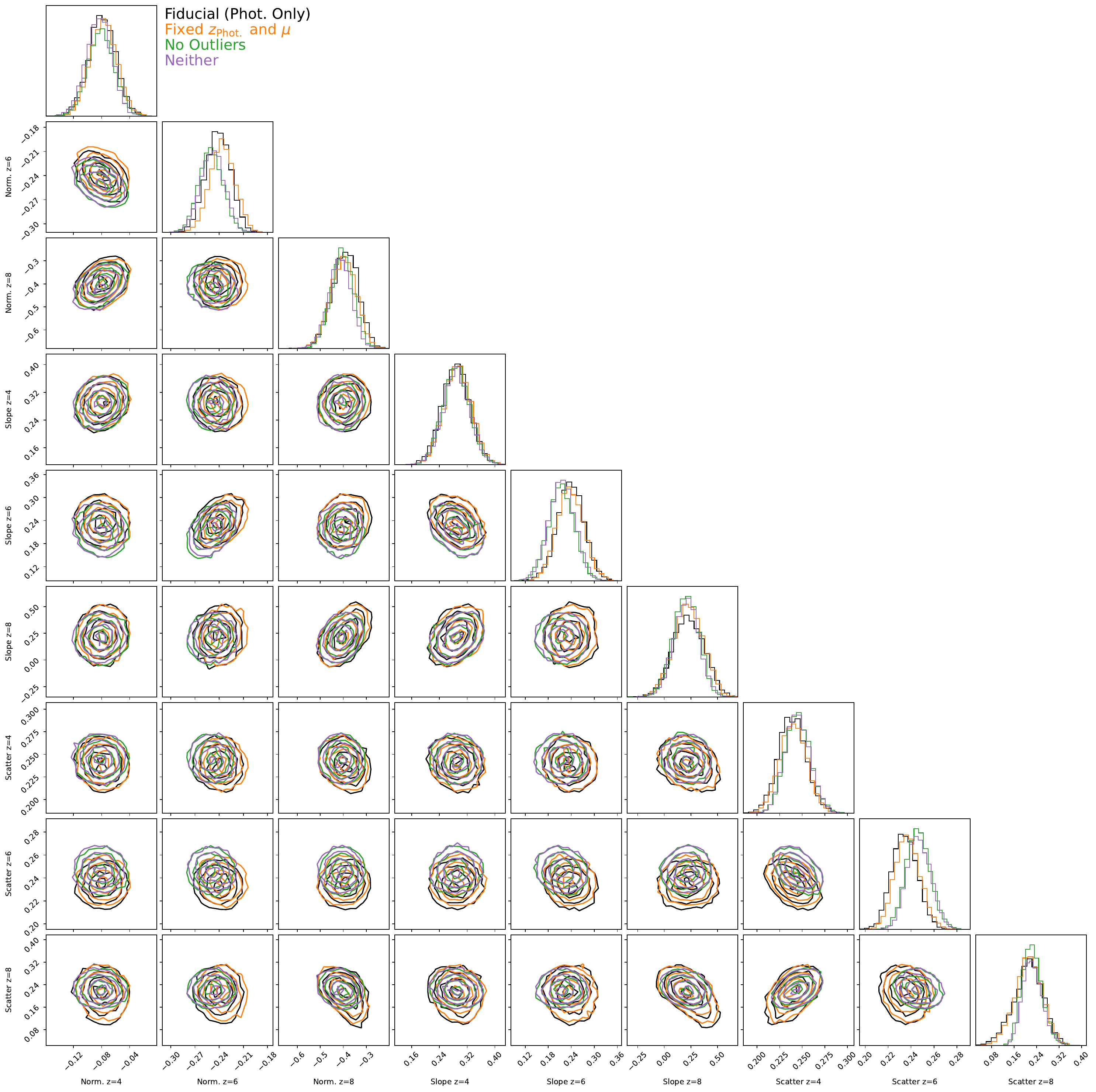}
    \caption{The posterior distribution for the inferred parameters of the size-mass relationship using the photometric sample only. The joint distributions of each parameter (intercept, slope and scatter) at each knot of the continuity model is shown in a corner plot. The contours correspond to the 0.5,1,1.5 and 2 $\sigma$ regions of each distribution.  We compare the inference with and without the fiducial treatment of outliers and/or photometric redshift uncertainties.}
    \label{fig:meth_comp_post_corner}
\end{figure*}

The full posterior distribution of the size-mass inference is shown in a corner plot in Figure~\ref{fig:spec_phot_post_corner}. We compare the posterior distributions inferred using the photometric and spectroscopic samples themselves, and the combination. Most of the constraining power is in the photometric sample, with only marginal gains in precision when the spectroscopic sample is introduced. In addition we have introduced novel additions to the size-mass fitting procedure, such as the treatment of outliers and photometric redshift uncertainties. In Figure~\ref{fig:meth_comp_post_corner} we test how the addition of these methods affects the measured size-mass relationship. In general we-find very little difference between any of the variations. When the outlier treatment is not employed, the measured scatter appears to shift to higher values.

The consistency of the size-mass inference across all of these tests is a testament to the 20 band NIRCam coverage produced with the combination of UNCOVER and MegaScience programs which produces precise estimates on the stellar masses and photometric redshifts with few outliers. In other scenarios, with more uncertain photometric redshift these differences in methodology may have more of a larger effect.

There are additional potential systematics that arise because we are using observations from a lensing field. Lensing tends to be biased towards compact sources, as extended sources will be further stretched potentially leading them to be undetectable. We avoid the strongest biases caused by strong lensing by limiting to moderate magnifications, $\mu<4$ for our analysis sample. However there may still be systematics as we need to use the lensing model to convert between the measured sizes and the intrinsic physical size.

\begin{figure*}
    \centering
    \includegraphics[width = \textwidth]{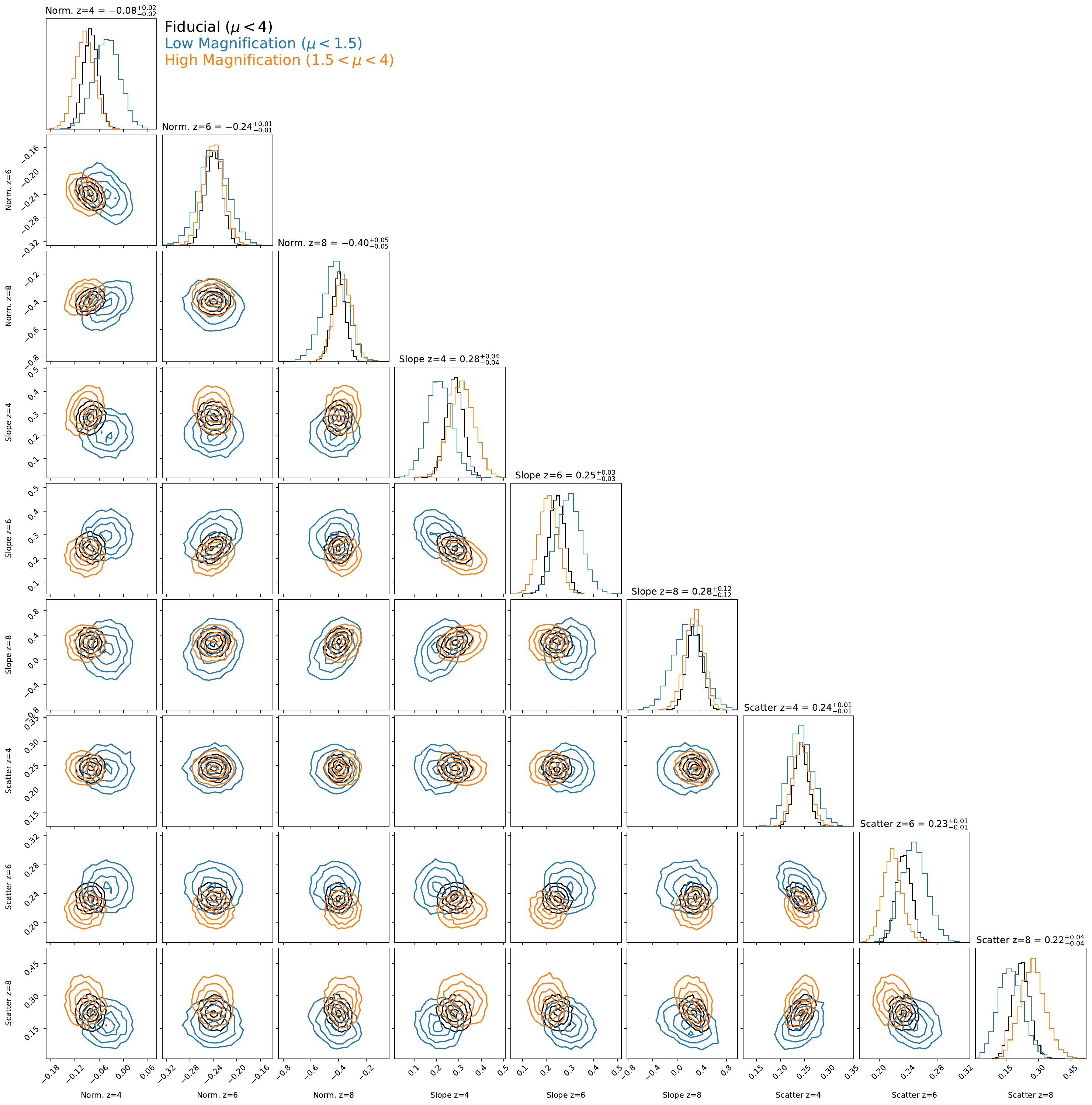}
    \caption{The posterior distribution for the inferred parameters of the size-mass relationship. The joint distributions of each parameter (Normalization, slope and scatter) at each knot of the continuity model is shown in a corner plot. The contours correspond to the 0.5,1,1.5 and 2 $\sigma$ regions of each distribution.  We compare the posterior from the fiducial sample to that with a lower magnification cut off, $mu <1.5$ to test for lensing related systematics.}
    \label{fig:mu_comp_post_corner}
\end{figure*}

We test for systematics related to lensing in Figure~\ref{fig:mu_comp_post_corner} by re-running the size-mass inference on a low magnification sample, $\mu<1.5$ and a high magnification sample, $\mu>1.5$. This low magnification sample contains 431 photometrically selected and 5 spectroscopically selected galaxies. Approximately half of the photometrically selected sample but the spectroscopic sample is biased towards high magnifications since the NIRSpec MSA was centered on the main cluster core~\citep{Price2024}. Additionally in Table~\ref{tab:size_mass_mu_comp} we displayed the mean and $1-\sigma$ standard deviation of the inferred parameters for the high $\mu$, low $\mu$ and fiducial sample. We find no systematic difference and all of the parameters agree within $1-\sigma$ when comparing to the fiducial sample, reassuring us that the results are not sensitive to any lensing related biases.

\begin{table}
    \centering
    \caption{\centering Inferred parameters for the optical size - stellar mass relationship comparing subsets of our sample separated by magnification. Additionally we include the number of photometrically and spectroscopically selected galaxies in each sample.}
    \begin{tabular}{c|ccc}
    & Fiducial ($\mu<4$) & Low $\mu$ ($\mu<1.5$) & High $\mu$ ($1.5<\mu<4$) \\ \hline
    $N_{\rm phot.} $ & 995 & 431  & 564 \\
    $N_{\rm spec.} $ & 35 & 5 & 30 \\ \hline
    Norm. ($z=4$) &  $-0.081 \pm 0.018$ &  $-0.041 \pm 0.032$ &  $-0.096 \pm 0.024$ \\
    Norm. ($z=6$) &  $-0.239 \pm 0.014$ &  $-0.239 \pm 0.025$ &  $-0.240 \pm 0.018$ \\
    Norm. ($z=8$) &  $-0.400 \pm 0.050$ &  $-0.425 \pm 0.092$ &  $-0.367 \pm 0.066$ \\ \hline
    Slope ($z=4)$ &  $0.281 \pm 0.038$ &  $0.214 \pm 0.055$ &  $0.314 \pm 0.055$ \\
    Slope ($z=6)$ &  $0.245 \pm 0.032$ &  $0.296 \pm 0.054$ &  $0.209 \pm 0.040$ \\
    Slope ($z=8)$ &  $0.276 \pm 0.120$ &  $0.169 \pm 0.246$ &  $0.273 \pm 0.153$ \\
    \hline
    Scatter ($z=4$) &  $0.244 \pm 0.015$ &  $0.243 \pm 0.026$ &  $0.244 \pm 0.018$ \\
    Scatter ($z=6$) &  $0.234 \pm 0.011$ &  $0.249 \pm 0.018$ &  $0.218 \pm 0.013$ \\
    Scatter ($z=8$) &  $0.220 \pm 0.040$ &  $0.172 \pm 0.058$ &  $0.276 \pm 0.059$ \\
    \end{tabular}
    \label{tab:size_mass_mu_comp}
\end{table}

\section{Gallery of Example Fits}
\label{sec:app_examps}

\begin{figure*}
    \centering
    \includegraphics[width = 0.48\textwidth]{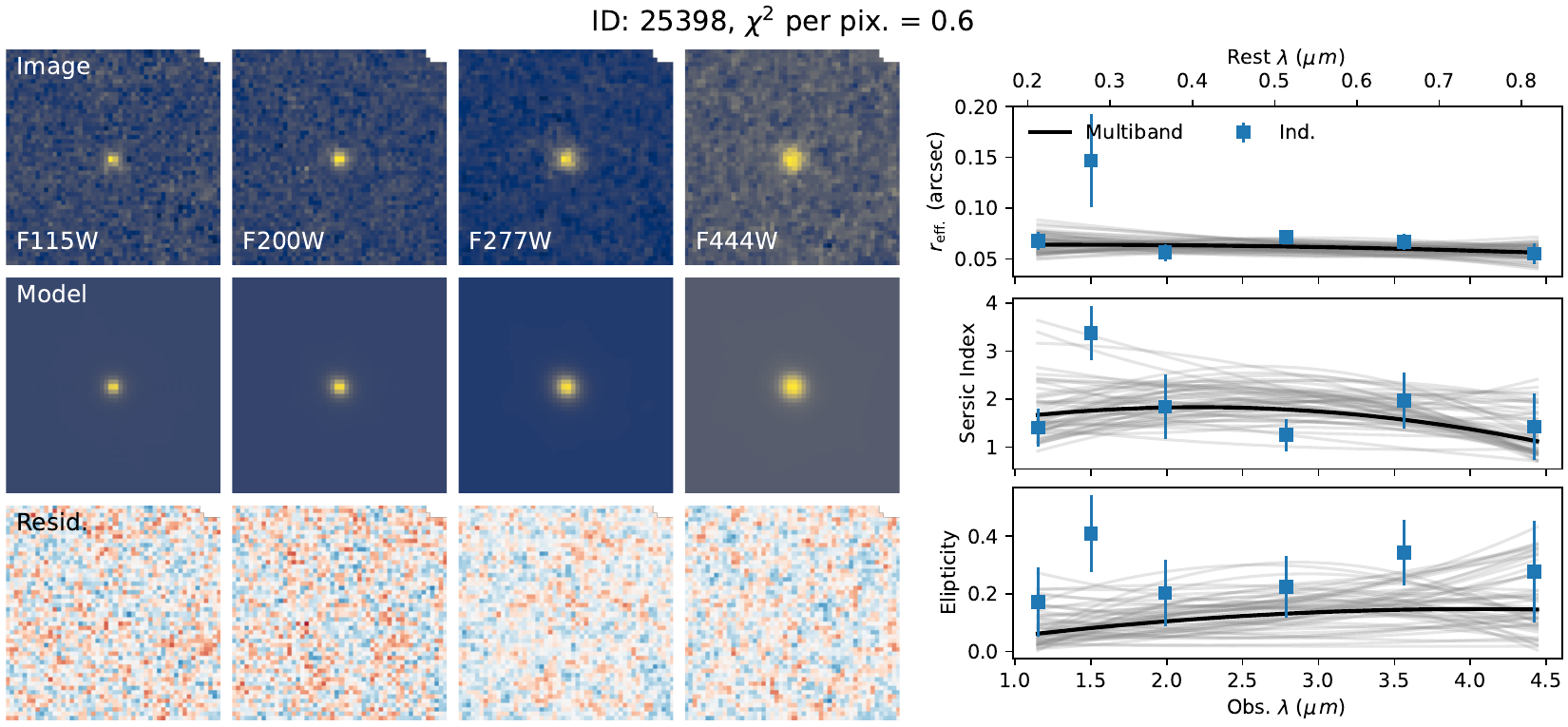}
    \includegraphics[width = 0.48\textwidth]{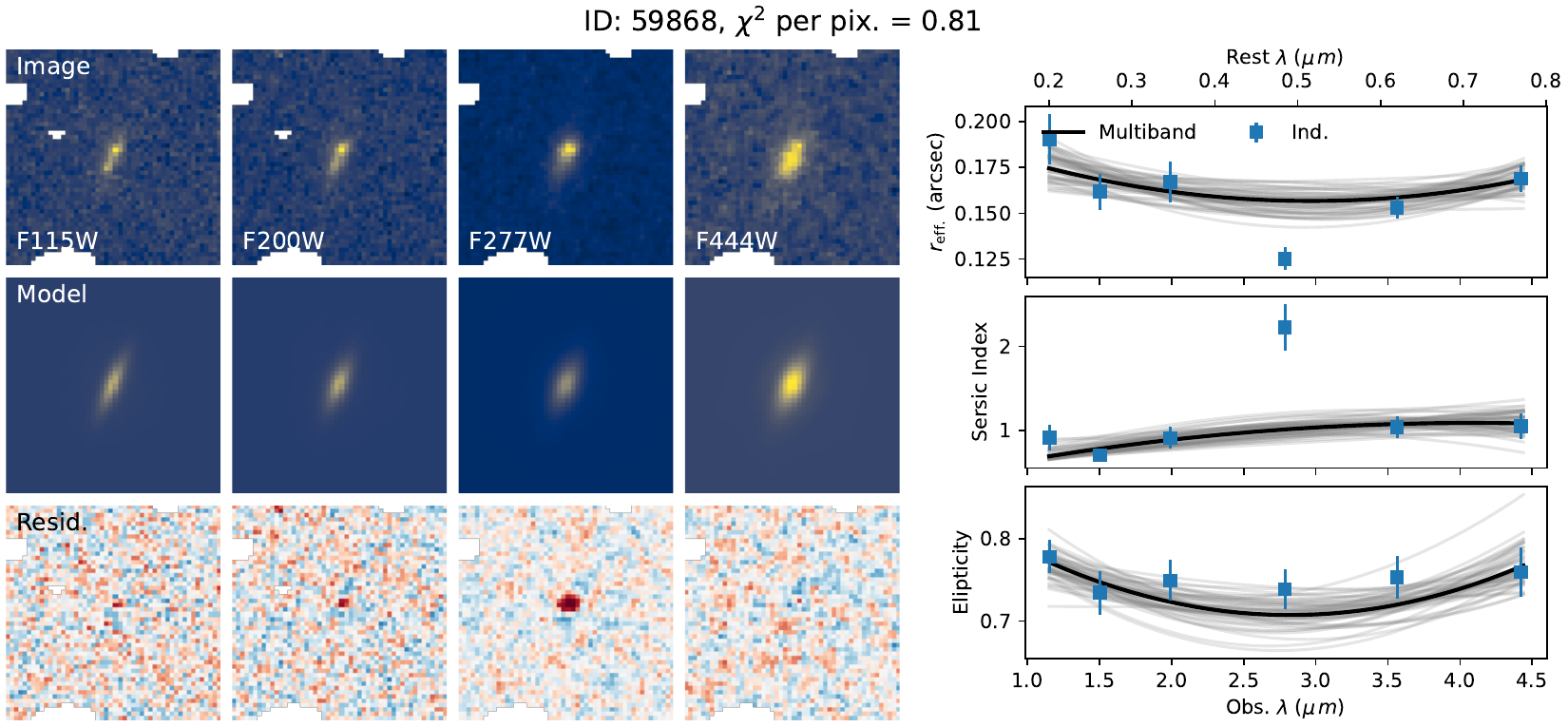}
    \includegraphics[width = 0.48\textwidth]{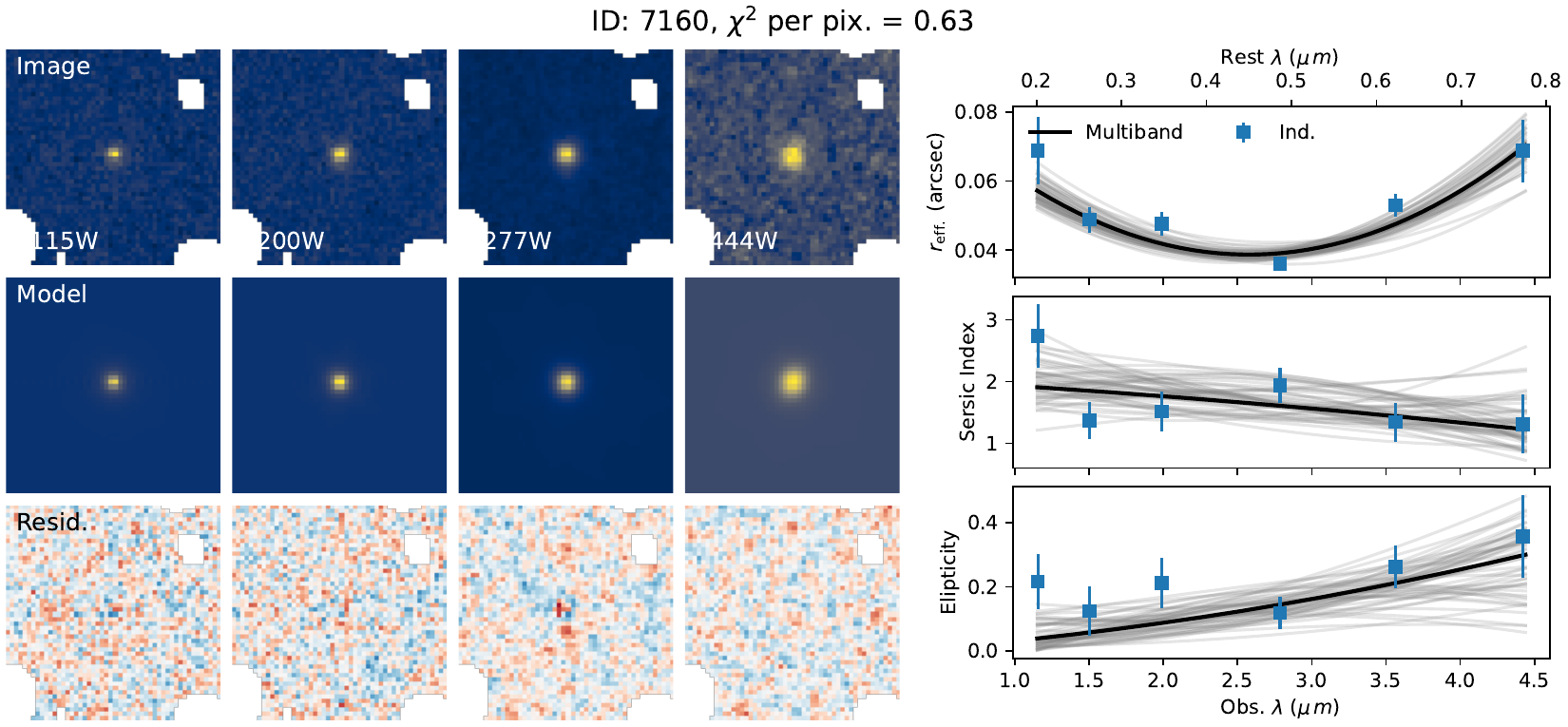}
    \caption{Further examples of the multi-band fits to galaxies similar to Fig.~\ref{fig:multi_showcase}. For each galaxy the images, best-fit models and residuals are shown in four bands (F115w,F200W, F277W and F444W) along with the wavelength dependent morphology comparing the multi-bands fits to those in individual bands. In this figure we highlight three examples galaxies with measured $\chi^2$ per pixel $<1.5$. These galaxies have smooth profiles that are well approximated by a Sersic profile}
    \label{fig:gallery_lowchi2}
\end{figure*}

\begin{figure*}
    \centering
    \includegraphics[width = 0.48\textwidth]{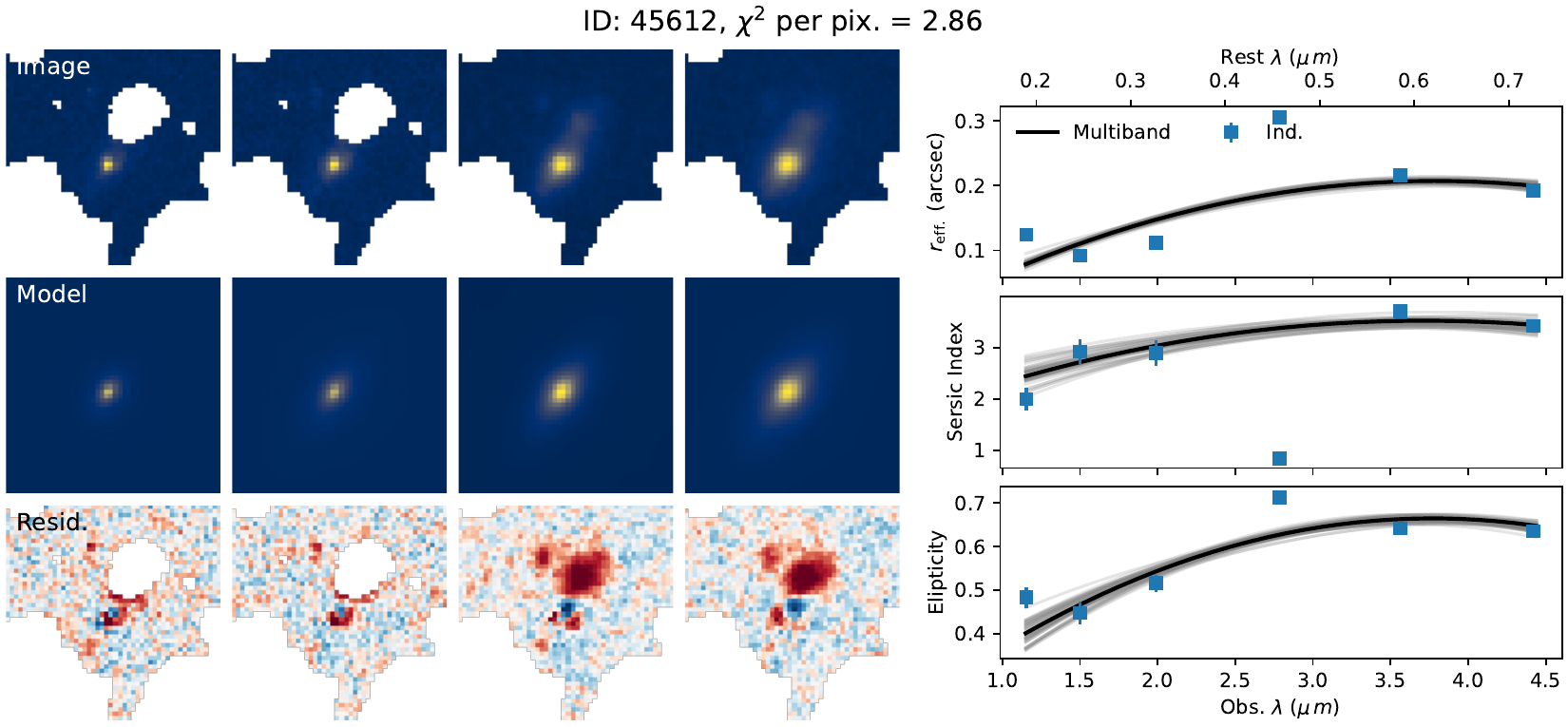}
    \includegraphics[width = 0.48\textwidth]{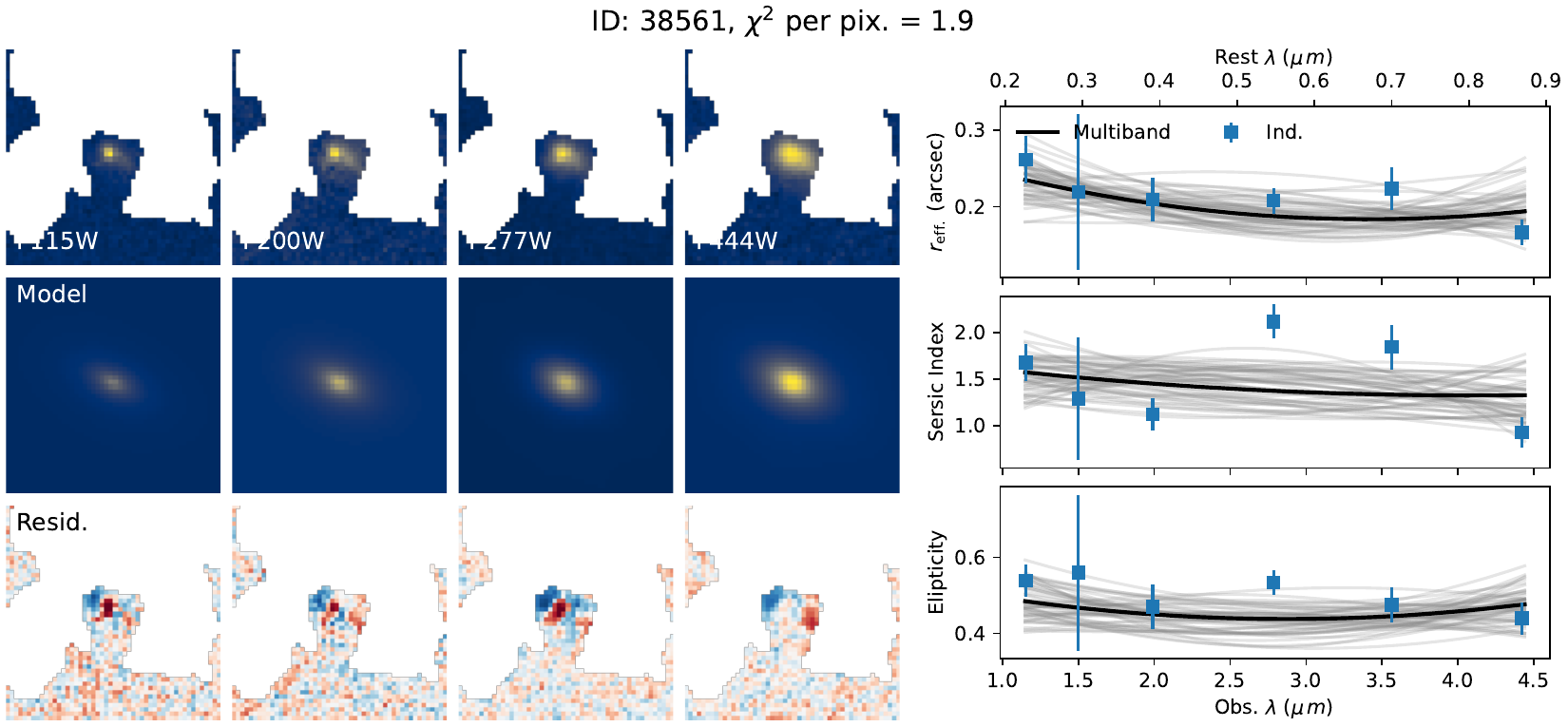}
    \includegraphics[width = 0.48\textwidth]{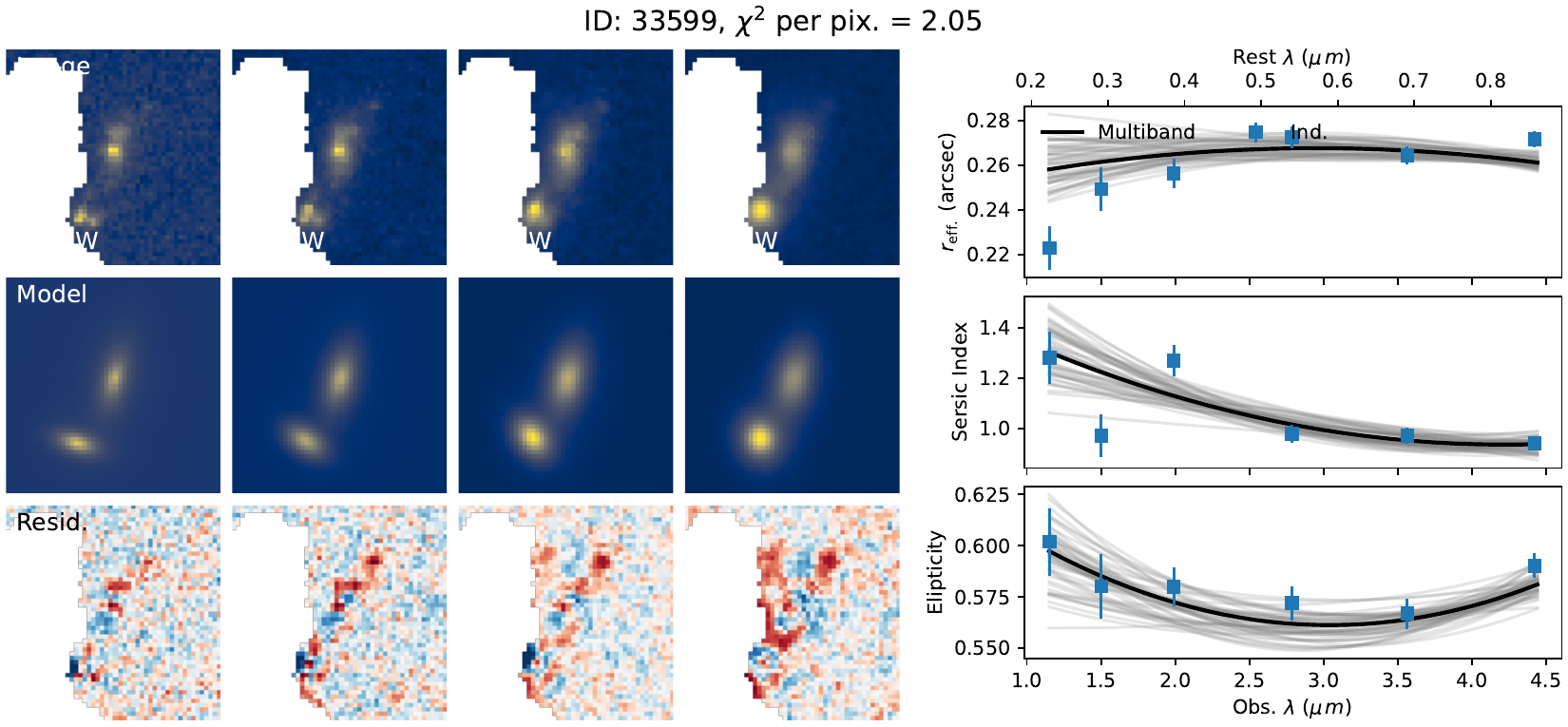}
    \caption{Further examples of the multi-band fits to galaxies similar to Fig.~\ref{fig:multi_showcase} and Fig.~\ref{fig:gallery_lowchi2}. In this figure we highlight three examples galaxies with measured $\chi^2$ per pixel between $1.5$ and $4$. These galaxies show some complex structure but their bulk light distribution is still well represented by a Sersic profile}
    \label{fig:gallery_midchi2}
\end{figure*}

\begin{figure*}
    \centering
    \includegraphics[width = 0.48\textwidth]{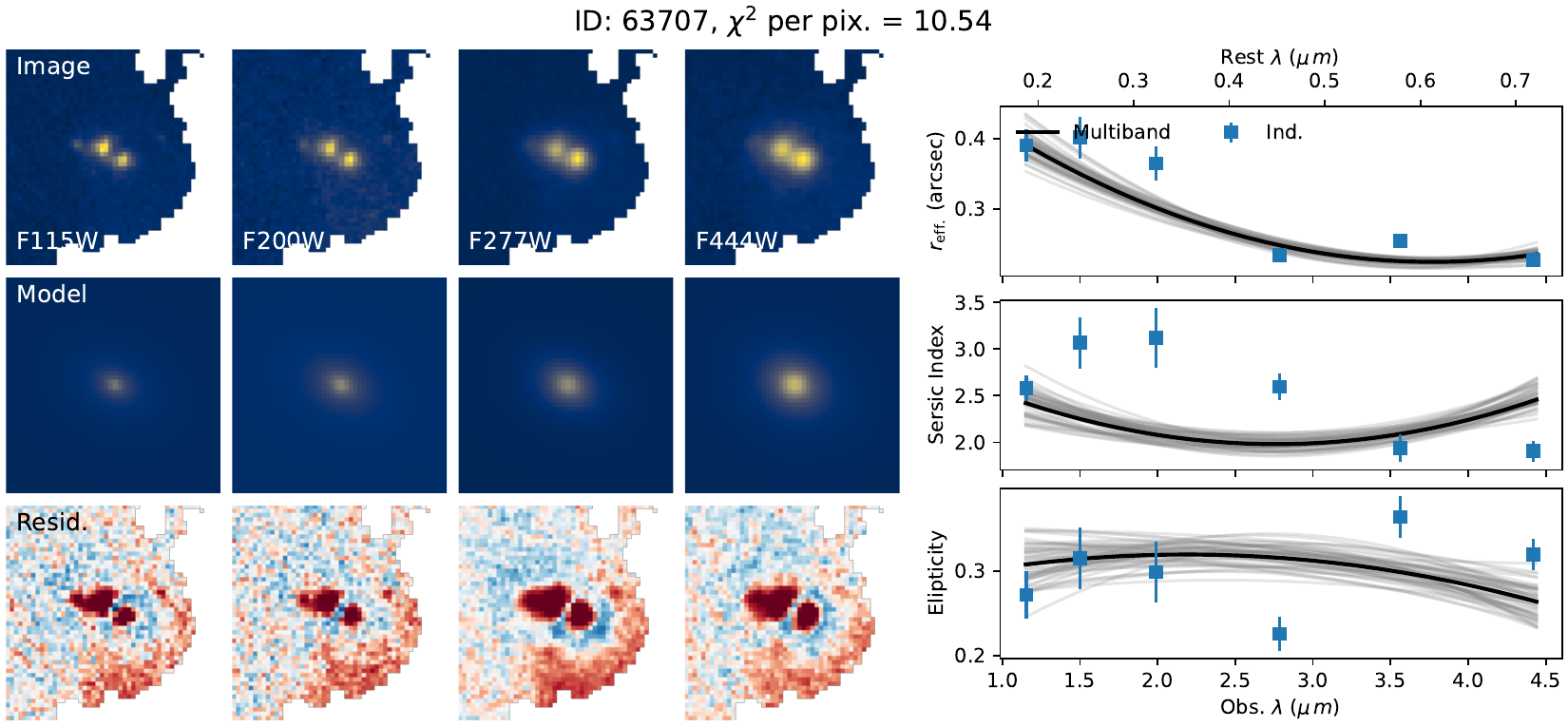}
    \includegraphics[width = 0.48\textwidth]{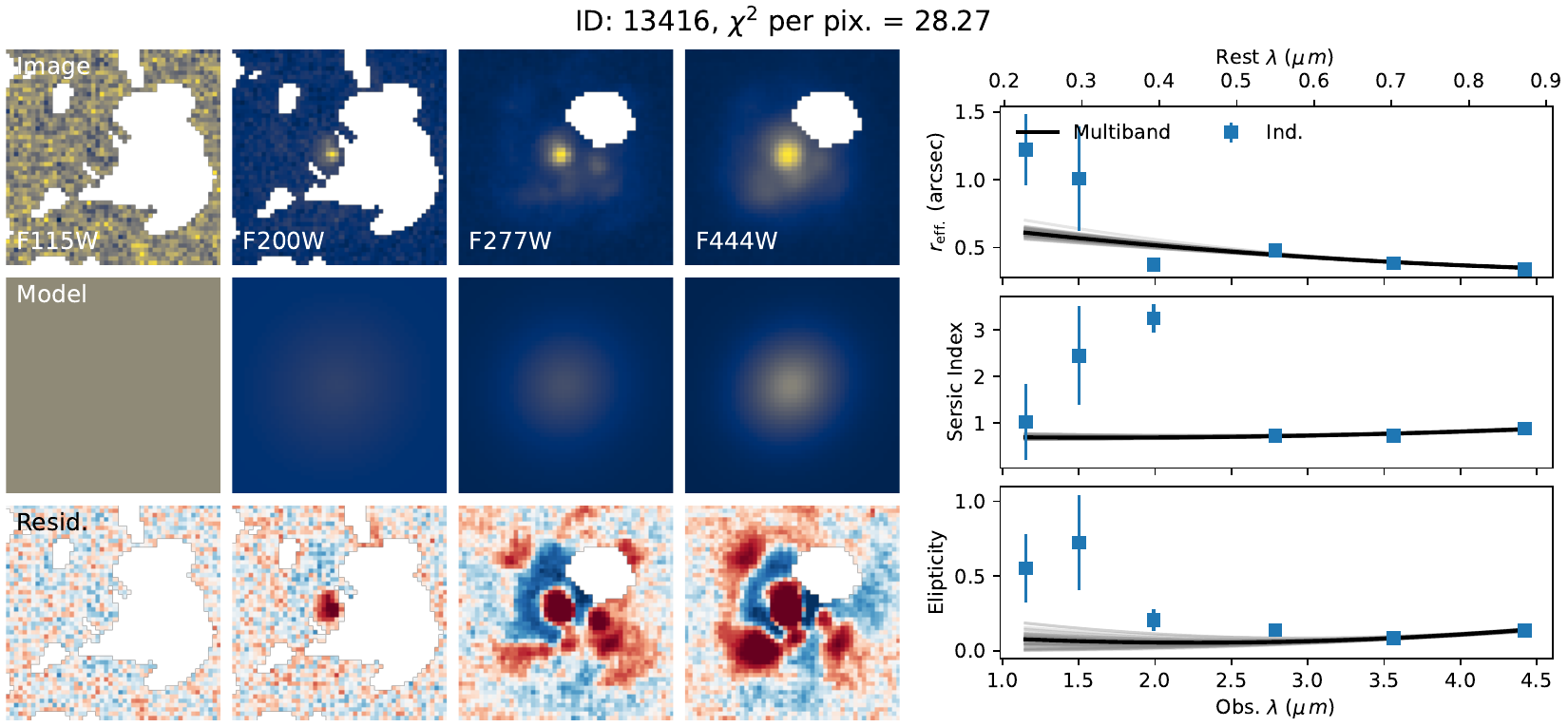}
    \includegraphics[width = 0.48\textwidth]{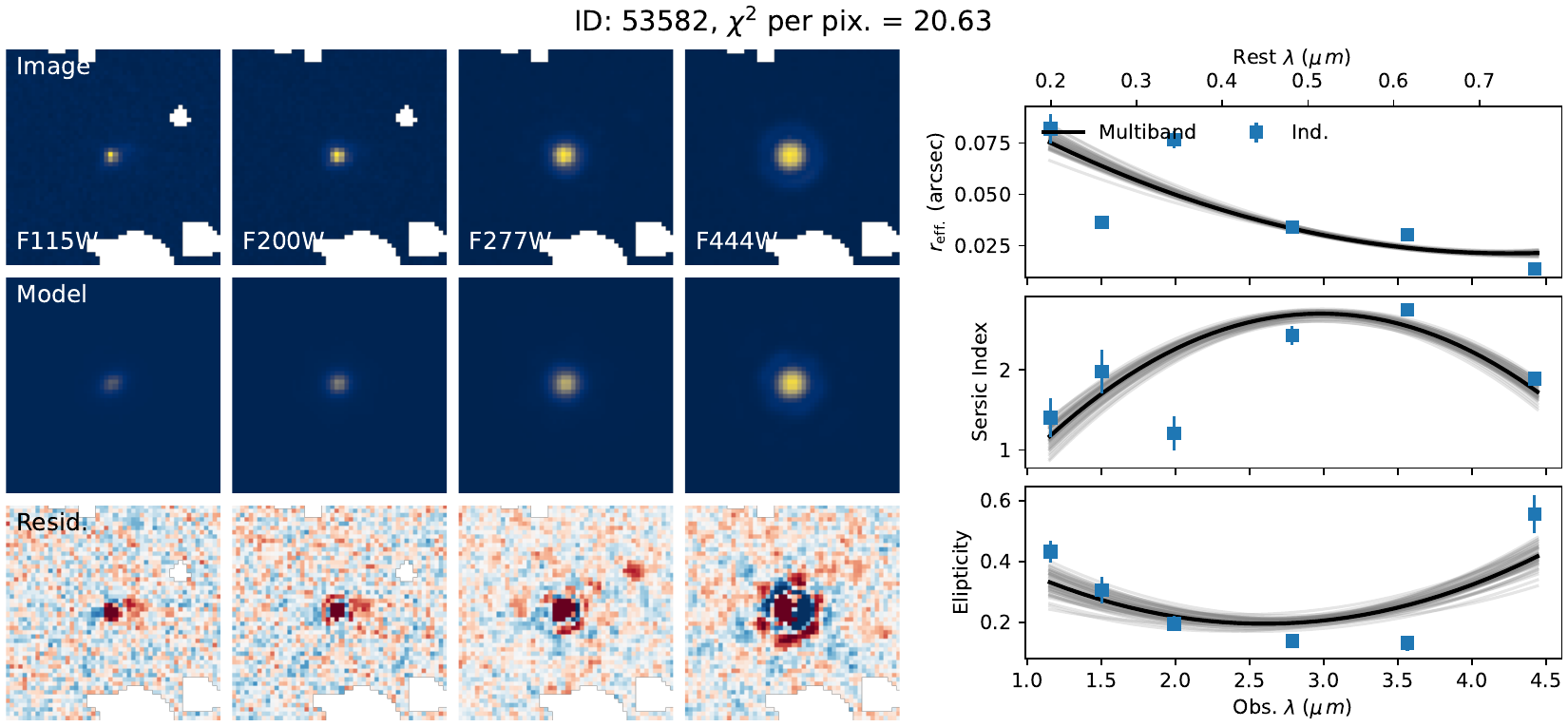}
    \caption{Further examples of the multi-band fits to galaxies similar to Fig.~\ref{fig:multi_showcase} and Fig.~\ref{fig:gallery_lowchi2}. In this figure we highlight three examples galaxies with measured $\chi^2$ per pixel $>4$ and are therefore excluded from the analysis sample. These galaxies are not well modeled by a Sersic profile, for example due to an obvious grouping of two sources into one or masking issues.}
    \label{fig:gallery_highchi2}
\end{figure*}

In this section we showcase more examples of the multi-band morphology fitting procedure. To showcase our choice of quality cuts we show three sets of galaxies in Figures ~\ref{fig:gallery_lowchi2}, ~\ref{fig:gallery_midchi2} and ~\ref{fig:gallery_highchi2}. These correspond to three slices in $\chi^2$ per pixel: $<1.5$, between $1.5$ and $4$ and $>4$, respectively. The latter of which are excluded from our analysis sample. The first set of galaxies have a smooth profile which is well modeled by a Sersic function. The middle set, with intermediate $\chi^2$ per pixel between 1.5 and 4, shows some complex structure but the bulk of the light profile appears well modeled by a Sersic function. The final set, which is cut from the analysis sample, showcases major issues in the fit, due to the grouping of two sources, masking issues or bright PSF sources which show deviations from the model PSF.

\begin{figure*}
    \centering
    \includegraphics[width = 0.48\textwidth]{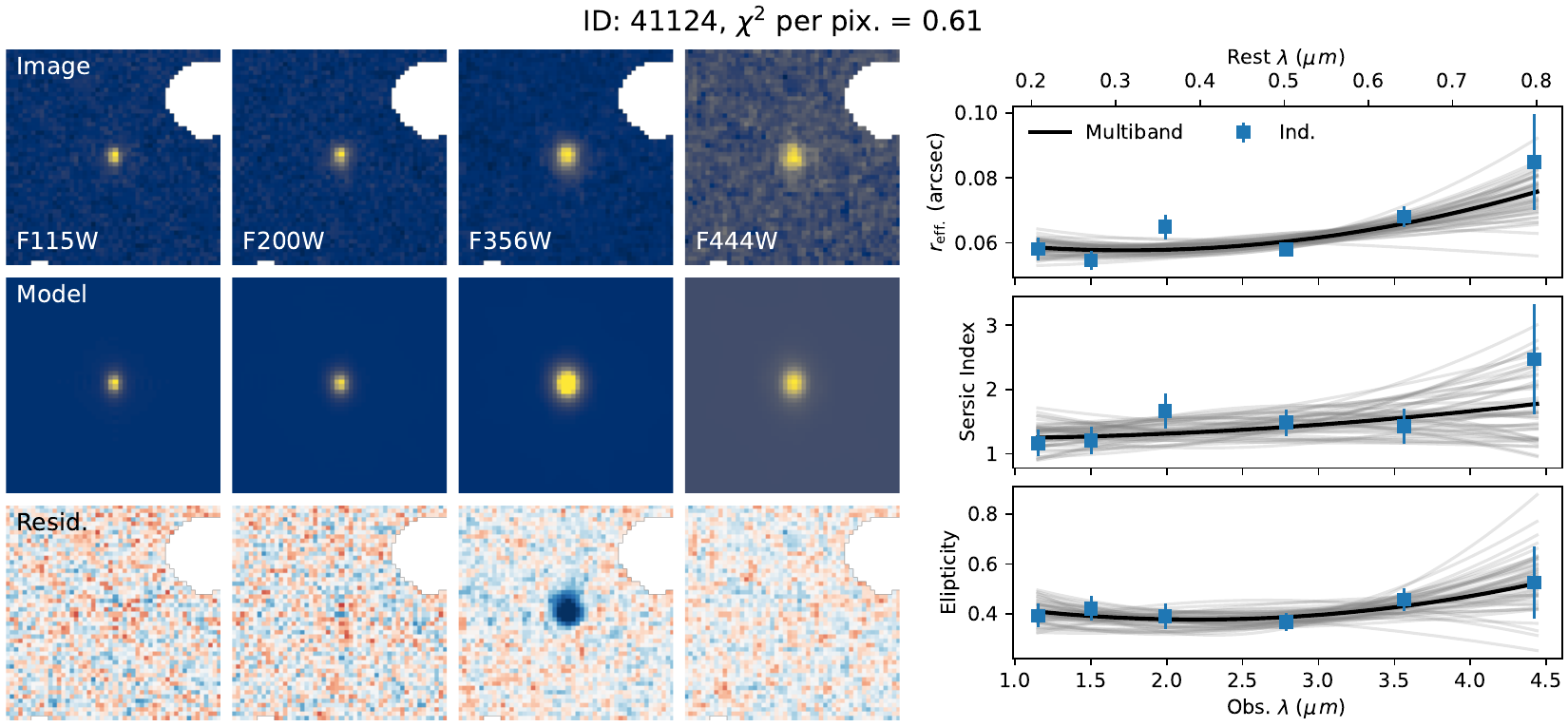}
    \includegraphics[width = 0.48\textwidth]{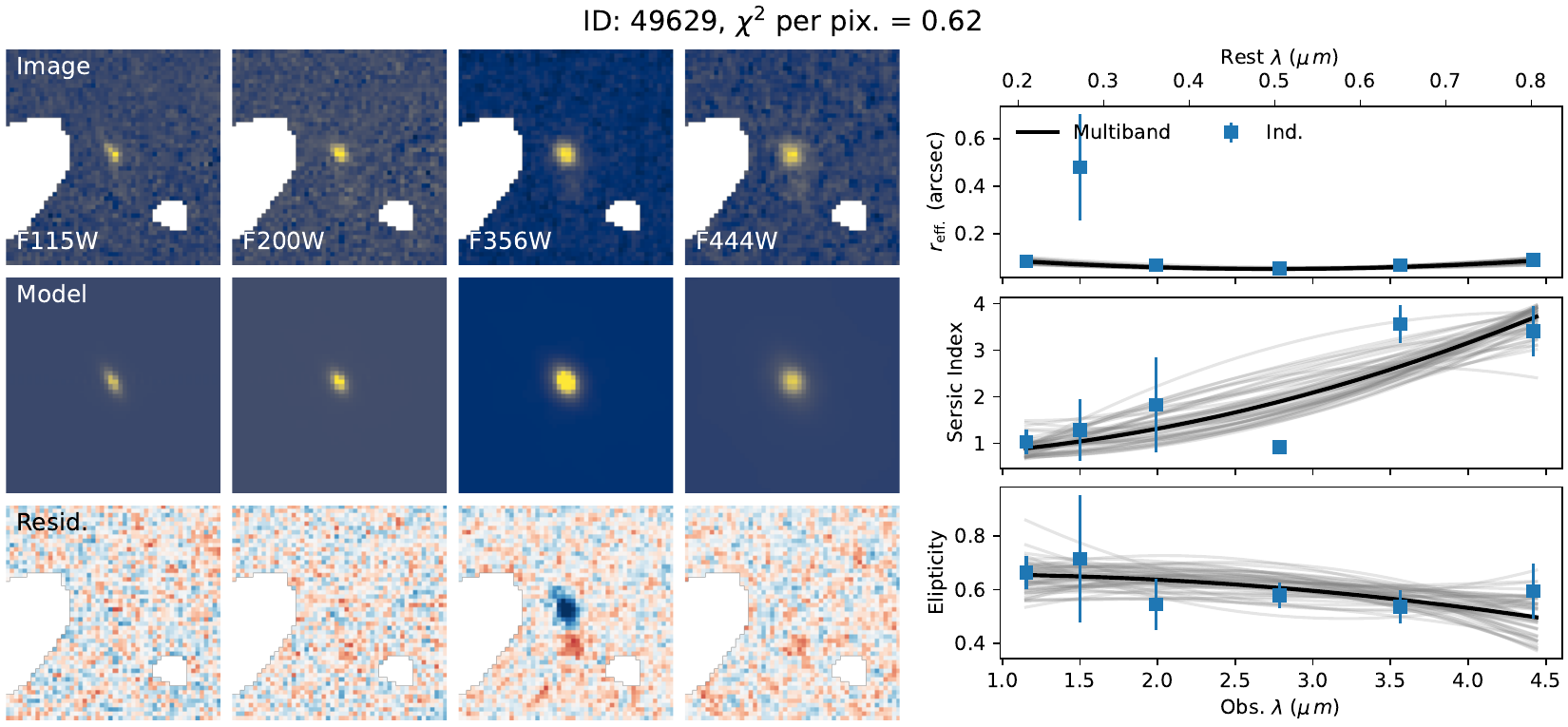}
    \includegraphics[width = 0.48\textwidth]{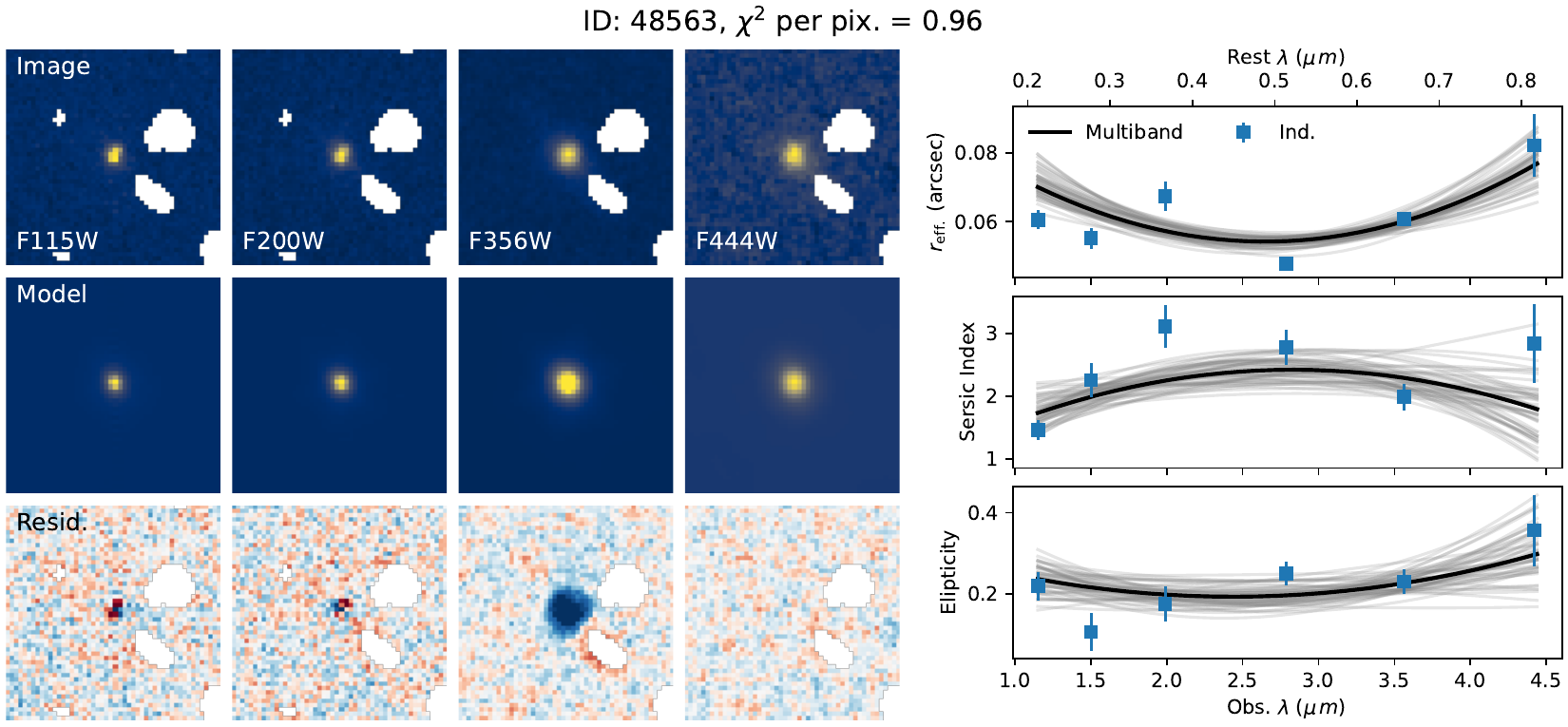}
    \caption{Further examples of the multi-band fits to galaxies similar to Fig.~\ref{fig:multi_showcase} and Fig.~\ref{fig:gallery_lowchi2}, except we are showing F356W as the third band instead of F277W. In this figure we highlight three examples galaxies where the broadband photometry is likely to be contaminated by line flux. To test the extreme case we select 3 galaxies at $4.3<z<4.6$ with \logmsol$<8.2$ and $SFR_{30} > 15\ M_\odot / {\rm yr}$. At this redshift the bright H$\alpha$-$[NII]$ complex is in F356W.}
    \label{fig:gallery_emline}
\end{figure*}

While we restrict our morphological analysis to broadband filters it is still possible for emission lines to dominate over the continuum, especially for the lowest mass galaxies in our sample. To investigate this we show 3 galaxies in Fig~\ref{fig:gallery_emline} at $4.3<z<4.6$ that are low mass and highly star-forming (\logmsol$<8.2$ and $SFR_{30} > 15\ M_\odot / {\rm yr}$). At this redshift the bright H$\alpha$-$[NII]$ complex redshifts into the F356W filter. This ``contamination'' does not have a large effect on the measured morphology of the galaxy. This could be because the stars and ionized gas have a similar extent or because the continuum still dominates the light. Although there appear to be residuals in the center of each galaxy where the Sersic model is overestimating the flux. This would be consistent with the extended $H\alpha$ profiles seen in \citet{Matharu2024} but we leave a full study of the emission line morphology to future work. Although this emission line contamination does not seem to greatly affect the multi-band or individual band morphology fits, even in these extreme cases.

\bibliography{all}{}
\bibliographystyle{aasjournal}

%% This command is needed to show the entire author+affiliation list when
%% the collaboration and author truncation commands are used.  It has to
%% go at the end of the manuscript.
%\allauthors

%% Include this line if you are using the \added, \replaced, \deleted
%% commands to see a summary list of all changes at the end of the article.
%\listofchanges

\end{document}